\documentclass[a4paper, 12pt]{article}
\usepackage{bm}
\usepackage[dvips]{graphicx, psfrag}
\usepackage{amsmath}
\usepackage{braket}
\usepackage{color}
\usepackage{ulem}
\usepackage{overpic}
\begin{document}
\begin{center}
  \Large{A numerical approach for particle-vortex interactions based on volume-averaged equations}\\
\end{center}

\newcommand{\lb}[1]{\textcolor{red}{#1}}
\newcommand{\lba}[1]{\textcolor{blue}{#1}}
\newcommand{\lbb}[1]{\textcolor{green}{#1}}
\newcommand{\lbc}[1]{\textcolor{black}{#1}}

\setlength{\baselineskip}{18pt}

\begin{center}
{\normalsize
    Toshiaki Fukada $^{1}$,
    Walter Fornari $^{2}$,
    Luca Brandt $^{2}$,
    Shintaro Takeuchi $^{3}$ \\
    and
    Takeo Kajishima $^{3}$ }\\
\end{center}
\begin{center}
$^{1}$ Central Research Institute of Electric Power Industry, \\
2-6-1 Nagasaka, Yokosuka, Kanagawa 240-0196, Japan\\
$^{2}$ Linn\'{e} Flow Center and Swedish e-Science Research Center (SeRC),
KTH Mechanics,\\ SE-10044 Stockholm, Sweden \\
$^{3}$ Department of Mechanical Engineering, Osaka University, \\
2-1 Yamada-oka, Suita, Osaka 565-0871, Japan
\end{center}

\begin{abstract}
To study the dynamics of particles in turbulence when their sizes are comparable to the smallest eddies in the flow, 
the Kolmogorov length scale, efficient and accurate numerical models for the particle-fluid interaction are still missing.
Therefore, we here extend the treatment of the particle feedback 
on the fluid based on the volume-averaged fluid equations (VA simulation) in the previous study of the present authors,
by estimating the fluid force correlated with the disturbed flow. 
We validate the model against interface-resolved simulations using the immersed-boundary method.
Simulations of single particles show that the history effect is well captured by 
the present estimation method based on the disturbed flow.
Similarly, the simulation of the flow around a rotating particle demonstrates that 
the lift force is also well captured by the proposed method.
We also consider the interaction between non-negligible size particles and an array of Taylor-Green vortices. 
For density ratios $\rho_d/\rho_c\geq 10$, the results show that
the particle motion captured by the VA approach is closer to that of the fully-resolved simulations
than that obtained with a traditional two-way coupling simulation. The flow disturbance is also well represented by the VA simulation.
In particular, it is found that history effects enhance the curvature of the trajectory in vortices
and this enhancement increases with the particle size.
Furthermore, the flow field generated by a neighboring particle at distances of around ten particle diameters
significantly influences particle trajectories.
The computational cost of the VA simulation proposed here is 
considerably lower than that of the interface-resolved simulation. 
\end{abstract}

\begin{flushleft}
\end{flushleft}

\section{Introduction}
Interactions between particles and a turbulent flow are important in many industrial processes like
cyclone separators and pulverised coal combustion.
Many factors determine the fluid-particle interaction such as the flow configuration, the particle relaxation time (the Stokes number),
the role of flow inertia (Reynolds number), the importance of gravity (Froude number), the solid volume fraction and the mass fraction (the latter two related by the density ratio).
One critical factor is the ratio between the particle size and a typical length scale of the flow.
In pipe flows and free jets laden with particles (including bubbles and droplets), for example,
the turbulence intensity increases when the particle diameter $D$ is
larger than one-tenth of the integral length scale (Gore and Crowe, 1989). 
Experimental works with dilute suspensions, on the other hand, 
report
significant reductions of the turbulence intensity 
when the particle diameter is comparable to the turbulence Kolmogorov length scale $\eta$ (Kulick et al., 1994; Paris and Eaton, 2001; Hwang and Eaton, 2006).
To understand the mechanisms of the interaction between the phases, numerical simulations can be used to
capture both the turbulence structures and the particle motion.
In many numerical studies, however, the force on the particle is approximated and the feedback force on the fluid is 
either ignored (i.e., one-way coupling) or simplified (i.e., traditional two-way coupling) to a point-source.
Thus, the turbulence attenuation by particles of $D\sim\eta$ is not reasonably reproduced by traditional
two-way coupling simulations (Eaton, 2009; Schneiders et al., 2017).
On the other hand, fully-resolved simulations like those in Kempe and Fr\"{o}hlich (2014), Picano et al. (2015), Fornari et al. (2016) 
and Santarelli and Fr\"{o}hlich (2016) are 
still too expensive for configurations of practical interest, which justifies the need for better models.

Focusing on the particle motion, the mean particle settling velocity is influenced by the background turbulence (Nielsen, 1993).
As recently shown in the experimental study of Good et al. (2014),  the settling velocity of particles can both increase and decrease
when the particle diameter is slightly smaller than the Kolmogorov length scale.
Although one-way coupling simulations capture this trend qualitatively,
a quantitative difference is recognised even for very dilute cases.
For direct numerical simulations of turbulent flows,
a grid width $\Delta x$ of the order of the Kolmogorov length scale is necessary.
Therefore, for a simulation of a particle-laden flow of particle size $D\sim\eta$,
an appropriate two-way interaction model between the flow and the particle  is required.

\begin{figure}[!h]
\centering
 \psfrag{pt}[l][l][1]{Particle centre}
\begin{overpic}[scale=0.6,angle=-0]{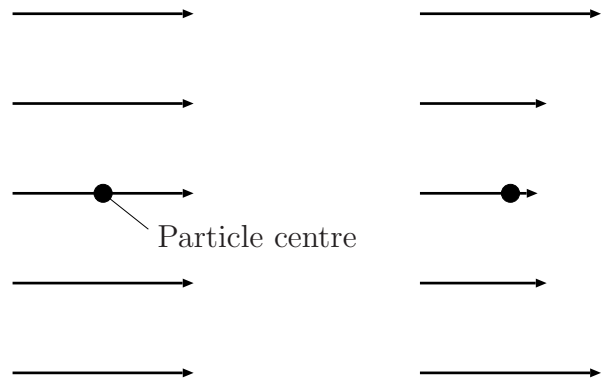}
\end{overpic}
\caption{Schematic images of undisturbed (left) and disturbed (right) velocity vectors
in a two-way coupling simulation around the particle centre.}
\label{fig:ud_vs_d}
\end{figure}

The most reliable numerical approach is resolving the particle boundary, in which case
the fluid force is directly computed. The immersed boundary method (IBM) is one of the possible approaches of this type as shown in several studies
(Kajishima and Takiguchi, 2002; Lucci et al., 2010; Tenneti and Subramaniam., 2014; Fornari et al., 2016).
As the particle diameter needs to be resolved by ten or more grid points, fully-resolved simulations
are practical when the particle is sufficiently larger than the Kolmogorov length scale. 
In other words, for the case $D\sim\eta$, fully-resolved simulations are not feasible
because of the prohibitive computational costs.
Therefore, the effect of the particle on the fluid has to be modelled without capturing the boundary layer.
In the traditional two-way coupling simulations, the drag force model is based on the undisturbed flow velocity and 
the particle is assumed to be much smaller than the Kolmogorov length and the grid width.
In the implementation, the local flow disturbance around the particle (see Fig.~\ref{fig:ud_vs_d}) is neglected and 
the disturbed velocity interpolated at the particle position is used as the undisturbed velocity in the expressions for the force
(Squires and Eaton, 1990; Boivin et al., 1998; Sundaram and Collins, 1999;
Li et al., 2001; Rani et al., 2004).
However, as Gualtieri et al. (2015) pointed out, the effect of the disturbance around the particle itself
cannot be ignored even when $D\ll\eta$.
These authors proposed, therefore, an estimation of the fluid force based on the Stokes flow around the particle, still considered to be smaller than the grid size.
In the case of $D\sim\eta$, the disturbance flow around the particle becomes more important and the assumption of Stokes flow is questionable.
On the fluid side, moreover, the point-source feedback force on the momentum equation is numerically distributed in space.
Since the particle size is ignored, the distribution does not consider the effect of the physical surface position.

To overcome these limitations, one possibility is volume averaging of the momentum equation that enable us to distribute
physically-meaningful feedback force.
This force is referred to as interaction force in this paper.
Fukada et al. (2016) recently developed a distribution model of the interaction force
for particle of diameters slightly larger than the grid size, $\Delta x$.
The interaction forces for uniform and simple shear flows around a sphere are modelled for particle Reynolds numbers ${\rm Re}_p=\mathcal{O}(10)$ and shear Reynolds number based on the particle diameter ${\rm Re}_{\gamma}=\mathcal{O}(1)$. The asymmetric distribution of the interaction force 
resulted in qualitatively and quantitatively reasonable flow fields consistent with the fully-resolved results.
The energy transfer on the volume-averaged field was also captured,
something which is not considered in traditional two-way coupling models.
However, the simulations in this previous work were limited to the case of a  fixed particle and known steady undisturbed flow.

In the present study, we therefore propose a novel estimation method of the fluid force based on the disturbed flow around the particle.
This approach is suited for the volume-averaged framework unlike a conventional two-way coupling approach. 
The study aims to show the applicability of the volume-average framework for the flow including moving particles.

We will initially consider the history effect on the particle motion, an effect whose importance is increasingly recognised 
(Olivieri et al., 2014; Daitche, 2015).
The history effects are highly influenced by the background flow and the modelling is therefore difficult
(Bagchi and Balachandar, 2003).
The traditional Basset history model based on the assumption of Stokes flow (Maxey and Riley, 1983)
is not applicable for a long physical time
since the model overestimates the past effects (Mei and Adrian, 1992).
Some models developed for finite Reynolds number are, on the other hand, limited to specific and relatively simple flows 
(Mei and Adrian, 1992; Wakaba and Balachandar, 2005).
The high computational cost of the integration of the history effect is also a factor to consider.
However, in an appropriate two-way coupling simulation, 
the history effects are included in the force estimation if the effect of unsteady disturbances is correctly captured (Gualtieri et al., 2015).
In a similar way, the lift force can be also represented by an appropriate two-way coupling algorithm that captures the flow disturbance, 
again reducing the dependence on a specific model.
To investigate how the history and the lift forces appear in the present simulation framework based
on the volume-averaged equation (referred to as VA simulation),
the settling of a particle in a fluid at rest and the flow around a rotating particle will be examined. 
For comparison and validation, we refer to the result from a fully-resolved IBM simulation, which is carried out in this study,
and previous results (Rubinow and Keller, 1961; Kurose and Komori, 1999, Bagchi and Balachandar, 2002; Bluemink et al., 2010).

We will then focus on the interaction between particles and a cellular vortical flow, the Taylor-Green vortex.
The particle diameter is $\mathcal{O}(10)$ times smaller than the vortex and its size is therefore non-negligible.
The particle trajectory in the Taylor-Green vortex has been first investigated by Maxey (1987) in the one-way coupling regime.
For particles of non-negligible size, however, the particle-vortex interaction 
leads to the flow disturbance at scales larger than the particle size as well as local disturbances around the particle.
Bergougnoux et al. (2014) showed in their experimental study that weak disturbances of the vortex influence the particle trajectory significantly.
Therefore, a two-way coupling investigation is necessary to correctly capture the particle motion in vortices.
Flow disturbances also induce and modify the interactions between two different particles.
The hydrodynamic forces on particles at distances of the order of $D$ in a uniform flow
have been the main objectives of previous studies
(Tsuji et al., 2003; Yoon and Yang, 2007; Ozgoren, 2013).
However, for a better understanding of the particle motion in turbulence,
inter-particle interactions at larger distances in a vortical flow should also be considered.
In this study, we simulate the behaviour of the particle in the Taylor-Green vortices
and show advantage of the VA approach for the particle motion in comparison to 
the point-model based method.
The fully-resolved IBM simulations are also carried out as reference.
Finally, the importance of two-way coupling on the particle trajectories is confirmed for cases with
different initial positions and inter-particle distance of around $10D$.

\section{Governing equations}

\subsection{Volume-averaged equations of the fluid phase}
The volume-averaged mass and momentum equations for dispersed multiphase flows 
are derived by Anderson and Jackson (1967)
under averaging length scales much larger than the inter-particle spacing.
The derivation is also detailed in Crowe et al. (1997).
On the other hand, the treatments of the residual stress and the interaction force terms 
for a case of averaging length scale comparable to the particle size have been developed by Fukada et al. (2016).
The volume-averaged equations and the models of these terms are briefly described in the following.

\begin{figure}[!h]
\centering
\begin{overpic}[scale=0.7,angle=-0]{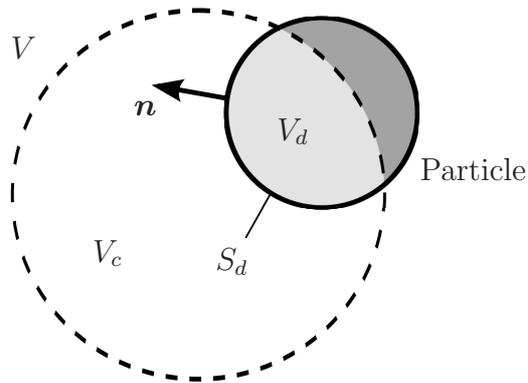}
  \put(0,80){$V$}
  \put(20,30){$V_c$}
  \put(65,60){$V_d$}
  \put(50,28){$S_d$}
  \put(30,66){$\bm{n}$}
  \put(100,50){Particle}
\end{overpic}
\caption{Schematic image of the volume-averaging area. 
Particle surface within $V$ is denoted by $S_d$.}
\label{fig:vol}
\end{figure}

The multiphase flow consists of the continuous (fluid) phase ($c$-phase) and the dispersed phase ($d$-phase).
Only rigid spherical particles are considered for the dispersed phase.
Considering the spherical averaging volume $V$ as shown in Fig.~\ref{fig:vol},
the volume fraction and the phase average of a physical property $B$ are defined as
\begin{eqnarray}
\alpha_k&=&\frac{V_k}{V},\label{eq:volume-fraction01}\\
\left<B\right>_k&=&\frac{1}{V_k}\int_{V_k}BdV,\label{eq:volume-average01}
\end{eqnarray}
where $V_k$ is the volume occupied by $k$-phase (either $c$ or $d$) inside $V$.
For a quantity $f$ defined through the averaging volume $V$,
we use the notation $f(\bm{x})$ where $\bm{x}$ denotes the centre of $V$.

The basic mass and momentum equations for an incompressible Newtonian fluid are written as follows:
\begin{eqnarray}
\nabla\cdot\bm{u}=0,
\label{eq:continuity01}
\end{eqnarray}
\begin{eqnarray}
\frac{\partial\bm{u}}{\partial t}+\nabla\cdot(\bm{u}\bm{u})=
-\frac{1}{\rho_c}\nabla p + \nu\nabla^2\bm{u}+\bm{h},
\label{eq:NS01}
\end{eqnarray}
where $\bm{u}$ is the velocity, $t$ time, $\rho_c$ the fluid density,
$p$ pressure, $\nu$ viscosity and $\bm{h}$ an external forcing.
In the present study, $\bm{h}$ is given to keep the background flow steady.
As shown in Fukada et al. (2016), through volume-averaging these equations, we obtain 
\begin{eqnarray}
\nabla\cdot(\alpha_c\left<\bm{u}\right>_c+\alpha_d\left<\bm{v}\right>_d)=0,
\label{eq:VAcontinuity01}
\end{eqnarray}
\begin{eqnarray}
\frac{\partial(\alpha_c\left<\bm{u}\right>_c)}{\partial t}
+\nabla\cdot(\alpha_c\left<\bm{u}\right>_c\left<\bm{u}\right>_c)
=-\frac{1}{\rho_c}\nabla P +\nu\nabla^2(\alpha_c\left<\bm{u}\right>_c+\alpha_d\left<\bm{v}\right>_d)
+\alpha_c\left<\bm{h}\right>_c-\nabla\cdot\bm{\tau}+\frac{\bm{f}}{V},
\label{eq:VANS01}
\end{eqnarray}
where $\bm{v}$ is the velocity inside a particle, $P$ the scalar function corresponding to pressure, 
$\bm{\tau}$ the residual stress and $\bm{f}$ the interaction force.
The fluid variables used in the simulations are $\alpha_c\left<\bm{u}\right>_c$ and $P$.
Appendix~A shows the differentiability of the volume-averaged quantities.
The form of the viscous term is different from that in other volume-averaged equations 
(Anderson and Jackson, 1967; Wachem et al., 2001), 
and the decomposition into $\nu\nabla^2(\alpha_c\left<\bm{u}\right>_c)+\nu\nabla^2(\alpha_d\left<\bm{v}\right>_d)$ is not allowed.

The scalar function $P$ can be decomposed as $\alpha_c\left<p\right>_c+\sum_n\alpha_{d,n}\hat{p}_n$
where $\alpha_{d,n}$ and $\hat{p}_n$ are the volume fraction and the surface-mean pressure (over the entire surface)
of the $n$th particle.
In a numerical simulation, $P$ can be obtained without considering the decomposition.

The residual stress is defined as
\begin{eqnarray}
\bm{\tau}=\alpha_c\left<\delta\bm{u}\delta\bm{u}\right>_c,
\label{eq:resst00}
\end{eqnarray}
where $\delta\bm{u}=\bm{u}-\left<\bm{u}\right>_c$.
The original model by Fukada et al. (2016) is
\begin{eqnarray}
\bm{\tau}&=&\alpha_c^{-1/3}\frac{R^2}{5}
\left\{\frac{\partial(\alpha_c\left<\bm{u}\right>_c+\alpha_d\left<\bm{v}\right>_d)}{\partial x_m}
-\alpha_d\left<\frac{\partial\bm{v}}{\partial x_m}\right>_d\right\}\nonumber\\
&&\left\{\frac{\partial(\alpha_c\left<\bm{u}\right>_c+\alpha_d\left<\bm{v}\right>_d)}{\partial x_m}
-\alpha_d\left<\frac{\partial\bm{v}}{\partial x_m}\right>_d\right\},
\label{eq:resst01}
\end{eqnarray}
where $x_m$ is the $m$th component of the Cartesian coordinates and
the summation convention is applied for the subscript $m$.
The velocity gradient inside the rigid particle $\partial\bm{v}/\partial x_m$
corresponds to the angular velocity of the particle.

The interaction force is defined as
\begin{eqnarray}
\bm{f}=-\int_{S_d}\left\{-\frac{1}{\rho_c}\delta p\bm{n}
+\nu\left(\nabla\bm{u}+(\nabla\bm{u})^T\right)\cdot\bm{n}\right\}dS
\label{eq:defintf}
\end{eqnarray}
where $\delta p$ is the deviation from the surface-mean pressure $\hat{p}_n$ of the corresponding particle,
$\bm{n}$ the unit normal vector on the particle surface directed to the fluid-phase and
$S_d$ the partitioned particle surface area inside $V$ (see Fig.~\ref{fig:vol}).

\subsection{Interaction force models}\label{sec:intf_models}
According to Appendix B of our previous study (Fukada et al., 2016),
the independent interaction force models can be successfully superimposed
for steady flows with ${\rm Re}_p\leq 40$ and $\gamma D^2/\nu=\mathcal{O}(1)$,
where $\gamma$ is the shear rate.
In the present study, we thus assume superposition of the different contributions:
\begin{eqnarray}
\bm{f}=\bm{f}_{\rm unif}+\bm{f}_{\rm vg}+\bm{f}_{\rm pg}+\bm{f}_{\rm rot},
\label{eq:intf_all01}
\end{eqnarray}
where the contributions due to the relative velocity, the undisturbed velocity gradient,
the undisturbed pressure gradient and the particle rotation are represented as 
$\bm{f}_{\rm unif}$, $\bm{f}_{\rm vg}$, $\bm{f}_{\rm pg}$ and $\bm{f}_{\rm rot}$.
Each component is described in the following.

Introducing the particle Reynolds number as
\begin{eqnarray}
{\rm Re}_p=\frac{|\bm{U}_r|D}{\nu},
\label{eq:defrep}
\end{eqnarray}
where $\bm{U}_r$ is the relative velocity based on the undisturbed flow
and $D$ is the particle diameter, the interaction force corresponding to the relative velocity
is modelled as (Fukada et al., 2016)
\begin{eqnarray}
\bm{f}_{\rm unif}(\bm{x})=
\left\{-\frac{F_{\rm drag}}{\rho_c}\xi+3\chi\xi(1-\xi)\frac{\bm{x}-\bm{x}_p}{|\bm{x}-\bm{x}_p|}\cdot\bm{m}\right\}\bm{m}
-\chi\xi(1-\xi)\frac{\bm{x}-\bm{x}_p}{|\bm{x}-\bm{x}_p|}.
\label{eq:intf_unif01}
\end{eqnarray}
In the above expression, $\xi=S_d/\pi D^2$ denotes the normalised surface area (hereafter, referred to as surface fraction), 
$\bm{m}$ the unit vector in the direction of the relative velocity $\bm{U}_r$,
$\bm{x}_p$ the particle centre position,
$F_{\rm drag}$ the contribution of the relative velocity to the fluid force
\begin{eqnarray}
F_{\rm drag}&=&3\pi\nu^2\rho_c{\rm Re}_p(1+0.15{\rm Re}_p^{0.687}),
\label{eq:fdragnl}
\end{eqnarray}
and $\chi$ is a fitting coefficient
\begin{eqnarray}
\chi&=&0.225\pi\nu^2{\rm Re}_p^{1.687}(1+0.126{\rm Re}_p^{0.464}).
\end{eqnarray}
Note that $V^{-1}\int \bm{f}_{\rm unif}d\bm{x}=-\bm{F_{\rm drag}}/\rho_c$ holds,
which guarantees the total momentum conservation between the phases.
The interaction force (\ref{eq:intf_unif01}) is based on the theoretical result of the flow around the sphere for low ${\rm Re}_p$ (Proudman and Pearson, 1965).
The fitting coefficient $\chi$ is introduced for extending to higher ${\rm Re}_p$ according to our numerical result (Fukada et al., 2016).

For a uniform shear flow of undisturbed velocity $\bm{U}_{\rm ud}=\gamma (x_2-x_{p2})\bm{e}_1$ with ${\rm Re}_p=0$, 
the shear-induced interaction force is modelled as (Fukada et al., 2016)
\begin{eqnarray}
\bm{f}_{{\rm vg}\gamma}(\bm{x})=-\pi\nu \gamma D^2\xi(1-\xi)
\left(4\frac{x_2-x_{p2}}{|\bm{x}-\bm{x}_p|}\bm{e}_1
+\frac{x_1-x_{p1}}{|\bm{x}-\bm{x}_p|}\bm{e}_2\right),
\end{eqnarray}
where $\bm{e}_i$ is the unit basis vector in the $i$th direction.
The model for a general undisturbed velocity gradient can be written as the superposition
\begin{eqnarray}
\bm{f}_{\rm vg}(\bm{x})=-\pi\nu D^2\xi(1-\xi)
\left\{4\left(\frac{\bm{x}-\bm{x}_p}{|\bm{x}-\bm{x}_p|}\cdot\nabla\right)\bm{U}_{\rm ud}
+(\nabla\bm{U}_{\rm ud})\cdot\frac{\bm{x}-\bm{x}_p}{|\bm{x}-\bm{x}_p|}\right\}.
\label{eq:intf_vg01}
\end{eqnarray}
Appendix~B of this paper briefly summarises the derivation, 
and for more details the reader is referred to Fukada et al. (2016).

The pressure gradient and the added-mass forces on the particle, $\bm{F}_{\rm pg}$,
are modelled as
\begin{eqnarray}
\bm{F}_{\rm pg}=-\frac{3m_c}{2\rho_c}\nabla P_{\rm ud}-\frac{m_c}{2}\frac{d\bm{v}_p}{dt},
\label{eq:fpg01}
\end{eqnarray}
where $P_{\rm ud}$ is the undisturbed pressure and $\bm{v}_p$ the translating velocity of the particle,
$m_c=\pi\rho_cD^3/6$ the mass of the displaced fluid.
Assuming that this force corresponds to the additional surface pressure  
$-(3\bm{F}_{\rm pg}\cdot\bm{n}/\pi D^2)$ obtained for an inviscid uniform flow,
the interaction force can be shown to be (see Appendix~B)
\begin{eqnarray}
\bm{f}_{\rm pg}(\bm{x})=-\xi\left\{1-(1-\xi)(1-2\xi)\right\}\frac{\bm{F}_{\rm pg}}{\rho_c}
-3\xi(1-\xi)(1-2\xi)\left(\frac{\bm{F}_{\rm pg}}{\rho_c}\cdot\frac{\bm{x}-\bm{x}_p}{|\bm{x}-\bm{x}_p|}\right)
\frac{\bm{x}-\bm{x}_p}{|\bm{x}-\bm{x}_p|}.
\label{eq:intf_pg01}
\end{eqnarray}

Using the Stokes solution for a rotating sphere with angular velocity $\bm{\Omega}_p$,
the interaction force due to the particle rotation is obtained as 
\begin{eqnarray}
\bm{f}_{\rm rot}(\bm{x})=
3\pi\nu D^2\xi(1-\xi)\bm{\Omega}_p\times\frac{\bm{x}-\bm{x}_p}{|\bm{x}-\bm{x}_p|}.
\label{eq:intf_rot01}
\end{eqnarray}

\subsection{Estimation of undisturbed flow at the particle position}\label{sec:estimate_undisturbed_flow}
To represent the interaction between the fluid and the particles,
the relative velocity $\bm{U}_r$ and the undisturbed gradients of $\bm{U}_{\rm ud}$ and $P_{\rm ud}$
at the particle position need to be estimated.
In traditional two-way coupling simulations,
the disturbed fluid velocity interpolated at the particle position is often regarded 
as the undisturbed flow when computing the interaction.
However, this treatment is justified only when the particle is much smaller than the grid spacing
and not appropriate for the present cases ($D\sim\Delta x$).
Therefore, in this paper, we propose new estimation methods for the relative velocity $\bm{U}_r$
and the undisturbed gradients of $\bm{U}_{\rm ud}$ and $P_{\rm ud}$.
The radius $R$ of the averaging volume $V$ is kept to be $0.75D$ throughout this paper
as the effectiveness of this value was confirmed (Fukada et al., 2016).

\subsubsection{Relative velocity}\label{sec:estimate_drag}
\begin{figure}[!h]
\centering
 \psfrag{u}[s][s][1]{${\alpha_c|\left<\bm{u}\right>_c-\bm{v}_p|D}/{\nu}$}
 \psfrag{rep}[c][c][1]{${\rm Re}_p$}
\begin{overpic}[scale=0.6, angle=-90]{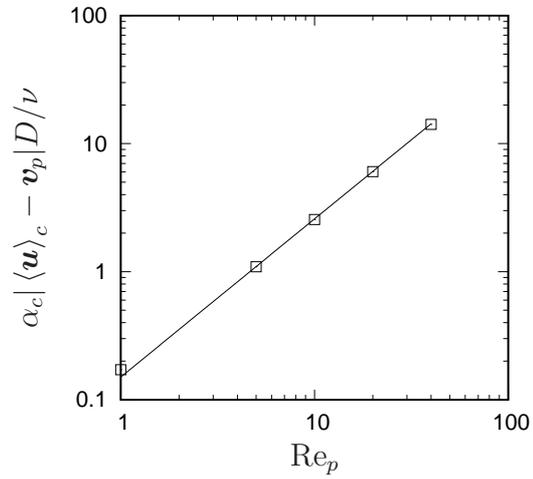}
\end{overpic}
\caption{Comparison of the averaged velocity 
at the particle centre from the numerical data (symbol) and from the model by Eq.\ (\ref{eq:rep_estimate01}) (line).}
\label{fig:u-rep-fit}
\end{figure}

The contribution of the relative velocity to the fluid force (\ref{eq:fdragnl}) is described in terms of ${\rm Re}_p$.
As the volume-averaged disturbed velocity $\alpha_c\left<\bm{u}\right>_c$ is obtained
from the volume-averaged equations, 
the correlation for the particle Reynolds number ${\rm Re}_p$ is necessarily based on $(\alpha_c\left<\bm{u}\right>_c)(\bm{x}_p)$.
For the fixed value $R=0.75D$, the correlation equation,
\begin{eqnarray}
{\rm Re}_p=4.64\left(\frac{\alpha_c|\left<\bm{u}\right>_c-\bm{v}_p|D}{\nu}\right)^{0.81},
\label{eq:rep_estimate01}
\end{eqnarray}
is obtained by a curve fitting based on the numerical data of our previous study (Fukada et al., 2016)
for the steady uniform flow around a single particle.
Figure~\ref{fig:u-rep-fit} shows that Eq.~(\ref{eq:rep_estimate01}) is valid for ${\rm Re}_p\leq 40$,
which is sufficient for the present study.
The effect of unsteady velocity on the force is qualitatively reflected through $|\left<\bm{u}\right>_c-\bm{v}_p|$.
The estimation method for $\alpha_c(\left<\bm{u}\right>_c-\bm{v}_p)$ is in Sec.~\ref{sec:va_scheme}.
In the numerical implementation, 
the direction of the relative velocity is assumed to be the same as 
that of $\alpha_c(\left<\bm{u}\right>_c-\bm{v}_p)$.
  
\subsubsection{Velocity and pressure gradients}\label{sec:estimate_vpgrad}
\begin{figure}[!h]
\centering
 \psfrag{du}[s][s][1]{$-\delta_{urr}D^2/\nu$}
 \psfrag{rep}[c][c][1]{${\rm Re}_p$}
\begin{overpic}[scale=0.6, angle=-90]{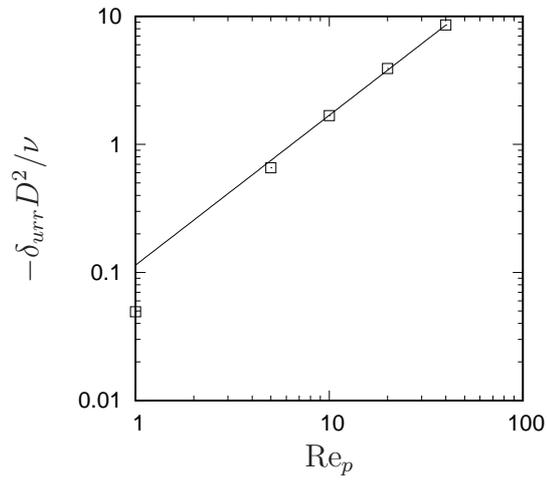}
\end{overpic}
\caption{Comparison of the disturbed velocity gradient at the particle centre 
from the numerical data (symbol) and from the model by Eq.\ (\ref{eq:fitvg_ave1}) (line).}
\label{fig:du-rep-fit}
\end{figure}

\begin{figure}[!h]
\centering
 \psfrag{dp}[s][s][1]{$-\delta_{pr}D^3/(\rho_c\nu^2)$}
 \psfrag{rep}[c][c][1]{${\rm Re}_p$}
\begin{overpic}[scale=0.6, angle=-90]{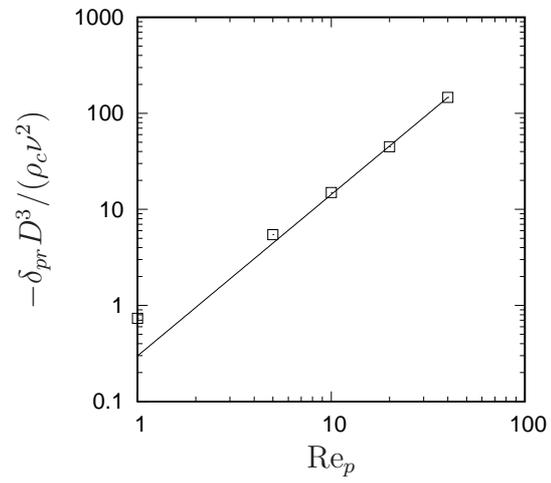}
\end{overpic}
\caption{Comparison of the disturbed pressure gradient at the particle centre
from the numerical data (symbol) and from the model by Eq.\ (\ref{eq:fitpg_ave1}) (line).}
\label{fig:dp-rep-fit}
\end{figure}

A simple estimation method for the undisturbed gradients based on the volume-averaged variables is proposed.
We define the differentiation operator in the direction $\bm{e}_i$ at the particle centre:
\begin{eqnarray}
\delta(f,l\bm{e}_i)=
\frac{f(\bm{x}_p+l\bm{e}_i)-f(\bm{x}_p-l\bm{e}_i)}{2l},
\end{eqnarray}
where $f$ is a function of $\bm{x}$,
and $l$ is an appropriately determined distance.
In this study, $l=D$ is adopted as supported by the tests 
for the pressure gradient shown in Appendix~C.
One of the simplest possible estimates of the undisturbed gradients are
\begin{eqnarray}
\frac{\partial \bm{U}_{\rm ud}}{\partial x_i}\approx\delta(\alpha_c\left<\bm{u}\right>_c+\alpha_d\left<\bm{v}\right>_d,l\bm{e}_i)
,\;\;\;\frac{\partial P_{\rm ud}}{\partial x_i}\approx\delta(P,l\bm{e}_i).
\end{eqnarray}
However, even in a uniform flow, 
non-zero gradients would be estimated around a particle due to the particle relative motion. 
To remove the effect of the relative motion, 
the following equations are constructed for the uniform flow case by curve fittings:
\begin{eqnarray}
\delta_{urr}({\rm Re}_p)&=&\delta(\alpha_c\left<\bm{u}\cdot\bm{m}\right>_c+\alpha_d\left<\bm{v}\cdot\bm{m}\right>_d,D\bm{m})
\nonumber\\ &=&-0.114\left(\frac{\nu}{D^2}\right){\rm Re}_p^{1.17},
\label{eq:fitvg_ave1}\\[12pt]
\delta_{pr}({\rm Re}_p)&=&\delta(P,D\bm{m})
\nonumber\\ &=&-0.298\left(\frac{\rho_c\nu^2}{D^3}\right){\rm Re}_p^{1.68},
\label{eq:fitpg_ave1}
\end{eqnarray}
where the direction of the relative velocity $\bm{m}$ is $(\left<\bm{u}\right>_c-\bm{v}_p)/|\left<\bm{u}\right>_c-\bm{v}_p|$.
Figures~\ref{fig:du-rep-fit} and \ref{fig:dp-rep-fit} compare Eqs.~(\ref{eq:fitvg_ave1}) and (\ref{eq:fitpg_ave1}) 
with the fully-resolved numerical data obtained on a body-fit coordinate system (Fukada et al., 2016).
According to these correlations, the undisturbed gradients are approximated by
\begin{eqnarray}
\nabla\bm{U}_{\rm ud}&\approx& \bm{e}_i \delta(\alpha_c\left<\bm{u}\right>_c+\alpha_d\left<\bm{v}\right>_d,l\bm{e}_i)
-(1.5\delta_{urr}\bm{m}\bm{m}-0.5\delta_{urr}\bm{I}),
\label{eq:vg_estimate01}\\
\nabla P_{\rm ud}&\approx& \delta(P,l\bm{e}_i)\bm{e}_i - \delta_{pr}\bm{m},
\label{eq:pg_estimate01}
\end{eqnarray}
where $\bm{I}$ is the identity tensor and the summation convention is applied for the subscript $i$.

\subsection{Equation of motion for the dispersed phase}\label{sec:eom_particle}
For finite Reynolds numbers, the force $\bm{F}$ on a particle can be modelled as
\begin{eqnarray}
\bm{F}=F_{\rm drag}\bm{m}+m_c\frac{D\bm{U}_{\rm ud}}{Dt}
+\frac{m_c}{2}\left(\frac{D\bm{U}_{\rm ud}}{Dt}-\frac{d\bm{v}_p}{dt}\right)
+\bm{F}_{\rm h}+(m_d-m_c)\bm{g}+\bm{F}_{\rm ext}
\label{eq:fpcl01}
\end{eqnarray}
(see Crowe et al., 1997), where $\bm{F}_{\rm h}$ is the history force,
$\bm{g}$ the gravitational acceleration, $\bm{F}_{\rm ext}$ the external force,
$m_d=\pi\rho_dD^3/6$ the particle mass. 
The first three terms on the right-hand side are the steady viscous force, 
the pressure gradient force and the added-mass force.
The term $\rho_c(D\bm{U}_{\rm ud}/Dt)$ is replaced by $-\nabla P_{\rm ud}$ because
the viscous force $\nu\nabla^2\bm{U}_{\rm ud}$ cancels with the external force $\bm{h}$ 
for the undisturbed flows considered in the present study. 
The external force on the particle $\bm{F}_{\rm ext}$ is the same as that on the displaced fluid 
$\rho_c\int_{V_p}\bm{h}dV$ where $V_p$ is the volume of the particle.
For the history force at a finite Reynolds number,
a reliable model for general flows is not available so far.
However, as discussed by Gualtieri et al. (2015), 
the history effect can be partly reproduced in accurate two-way coupling simulations
without any specific model.
Therefore, in the present model, we will numerically solve the following equation
\begin{eqnarray}
\frac{d\bm{v}_p}{dt}=
\frac{2}{2m_d+m_c}\left\{F_{\rm drag}\bm{m}
-\frac{\pi D^3}{4}\nabla P_{\rm ud}+(m_d-m_c)\bm{g}+\rho_c\int_{V_p}\bm{h}dV\right\}.
\label{eq:vpcl01}
\end{eqnarray}
To investigate the effects of the pressure gradient and the added mass,
we will also compare the result by solving the following equation
\begin{eqnarray}
\frac{d\bm{v}_p}{dt}=
\frac{1}{m_d}\left\{F_{\rm drag}\bm{m}+(m_d-m_c)\bm{g}+\rho_c\int_{V_p}\bm{h}dV\right\}.
\label{eq:vpcl02}
\end{eqnarray}
The particle position is given by the following equation:
\begin{eqnarray}
\frac{d\bm{x}_p}{dt}=\bm{v}_p.
\label{eq:xpcl01}
\end{eqnarray}
To include a first-order approximation of the effect of the flow on the particle rotation,
the following equation for the angular velocity based on the Stokes solution is considered
\begin{eqnarray}
\frac{d\bm{\Omega}_p}{dt}=
\frac{\pi\rho_c\nu D^3}{I_d}\left(\frac{1}{2}\nabla\times\bm{U}_{\rm ud}-\bm{\Omega}_p\right),
\label{eq:opcl01}
\end{eqnarray}
where $I_d=m_dD^2/10$ is the moment of inertia of the spherical particle. 
This equation is consistent with Eqs.~(\ref{eq:intf_vg01}) and (\ref{eq:intf_rot01})
as both equations assume the same Stokes solution.


\section{Numerical methods}
For simplicity, the simulation with the volume-averaged equations (Secs.~2.1 and 2.2)
is referred to as VA simulation.
To investigate the history effect, a one-way coupling simulation is also attempted.
Moreover, we will compare the results of the VA simulation to those of a traditional two-way coupling simulation
and the fully-resolved simulation with the immersed boundary method (IBM).
In the following, these numerical procedures are described briefly.
In all the simulations, 
the 2nd-order central-difference scheme is used for the spatial derivatives 
with a staggered arrangement for the fluid variables.
The fractional step method (Kim and Moin, 1985) is used as the pressure-velocity coupling algorithm for the fluid-phase.
The computational cell is a cube of side length $\Delta x$.

\subsection{VA simulation}\label{sec:va_scheme}
The 2nd-order Runge-Kutta method is used for the time evolution of both phases.
As the averaging volume $V$ is larger than the particle ($R=0.75D$),
the contribution of each particle to the volume fraction $\alpha_d$ is calculated as
\begin{eqnarray}
\alpha_d(\bm{x})=
   \begin{cases}
     \; \dfrac{r_d^3}{R^3}
     \;\;\;\;\;\;\;\;\;
         {\rm for~}\; 0\leq |\bm{x}-\bm{x}_p|<R-r_d \\[12pt]
     \; \dfrac{1}{16R^3} \left\{
     |\bm{x}-\bm{x}_p|^3
     -6(r_d^2+R^2)|\bm{x}-\bm{x}_p|
     -\dfrac{3(r_d^2-R^2)^2}{|\bm{x}-\bm{x}_p|}
     +8(R^3+r_d^3)\right\} \\[12pt]
     \;\;\;\;\;\;\;\;\;\;\;\;\;\;\;
         {\rm for~}\; R-r_d\leq |\bm{x}-\bm{x}_p|\leq R+r_d \\[12pt]
     \; 0
     \;\;\;\;\;\;\;\;\;\;\;\;
         {\rm for~}|\bm{x}-\bm{x}_p|>R+r_d
   \end{cases},
   \label{eq:volfrac_calc}
\end{eqnarray}
where $r_d=D/2$ is the radius of the particle.
The surface fraction $\xi$ is calculated as
\begin{eqnarray}
\xi(\bm{x})=
   \begin{cases}
     \; 1
     \;\;\;\;\;\;\;\;\;\;\;\;\;\;\;\;\;\;\;\;\;\;\;\;\;\;\;\;\;\;\;\;\;\;\;\;\;
     \;\;\;\;\;\;\;\;\;\;\;\;\;\;
         {\rm for~}\; 0\leq |\bm{x}-\bm{x}_p|<R-r_d \\[12pt]
     \; \dfrac{1}{2}\left(1-\dfrac{|\bm{x}-\bm{x}_p|}{2r_d}
     +\dfrac{R^2-r_d^2}{2r_d|\bm{x}-\bm{x}_p|}\right)
     \;\;
         {\rm for~}\; R-r_d\leq |\bm{x}-\bm{x}_p|\leq R+r_d \\[12pt]
     \; 0
     \;\;\;\;\;\;\;\;\;\;\;\;\;\;\;\;\;\;\;\;\;\;\;\;\;\;\;\;\;\;\;\;\;\;\;\;\;
     \;\;\;\;\;\;\;\;\;\;\;\;\;\;
         {\rm for~}|\bm{x}-\bm{x}_p|>R+r_d
   \end{cases}.
   \label{eq:surffrac_calc}
\end{eqnarray}
The averaged velocity of the solid-phase $\alpha_d\left<\bm{v}\right>_d$ is calculated as
\begin{eqnarray}
(\alpha_d\left<\bm{v}\right>_d)(\bm{x})=
   \begin{cases}
     \; \dfrac{r_d^3}{R^3}\bm{v}_p
     \;\;\;\;\;\;\;\;\;\;\;\;\;\;\;\;\;\;\;\;\;\;\;\;\;\;\;\;\;\;\;\;\;\;\;\;\;
     \;\;\;\;\;\;\;\;\;\;\;\;\;\;\;\;
         {\rm for~}\; 0\leq |\bm{x}-\bm{x}_p|<R-r_d \\[12pt]
     \;  \alpha_d(\bm{x})\bm{v}_p+K_d(|\bm{x}-\bm{x}_p|)\bm{\Omega}_p\times(\bm{x}-\bm{x}_p)
     \;\;\;
         {\rm for~}\; R-r_d\leq |\bm{x}-\bm{x}_p|\leq R+r_d \\[12pt]
     \; 0
     \;\;\;\;\;\;\;\;\;\;\;\;\;\;\;\;\;\;\;\;\;\;\;\;\;\;\;\;\;\;\;\;\;\;\;\;\;
     \;\;\;\;\;\;\;\;\;\;\;\;\;\;\;\;\;\;\;\;\;\;\;
         {\rm for~}|\bm{x}-\bm{x}_p|>R+r_d
   \end{cases},
   \label{eq:volgfrac_calc}
\end{eqnarray}
where the function $K_d(y)$ is 
\begin{eqnarray}
  K_d(y)=\dfrac{(R-r_d-y)^2(R+r_d-y)^2(R^2+4Ry-r^2+y^2)}{32R^3y^3}.
\end{eqnarray}
The derivation of these geometrical functions (\ref{eq:volfrac_calc})--(\ref{eq:volgfrac_calc}) is 
summarised in Appendix~D.

To keep the total external force on the system $\int\bm{h}dV$ constant,
the external forces are approximated as
\begin{eqnarray}
(\alpha_c\left<\bm{h}\right>_c)(\bm{x})&=&\alpha_c\bm{h}(\bm{x})
\end{eqnarray}
for fluid, in Eq.~(\ref{eq:VANS01}), and
\begin{eqnarray}
\rho_c\int_{V_p}\bm{h}dV=\sum_{ijk} \rho_c\alpha_{d,ijk}\bm{h}_{ijk}\Delta x^3
\end{eqnarray}
for solid, in Eq.~(\ref{eq:vpcl01}),
where the subscript $ijk$ represents the spatial point of grid index $(i,j,k)$.

The estimation of $\{\alpha_c(\left<\bm{u}\right>_c-\bm{v}_p)\}(\bm{x}_p)$ in Eq.~(\ref{eq:rep_estimate01})
is particularly important to predict the drag force.
As the distribution of $\alpha_c\left<\bm{u}\right>_c$ has a local minimum near the particle centre,
a linear interpolation is not sufficient.
Therefore, the following interpolation steps are used for the velocity:
\begin{eqnarray}
  \bm{w}&=&\alpha_c\left<\bm{u}\right>_c+\alpha_d\left<\bm{v}\right>_d,
  \label{eq:ipvelstep01}
  \\
  W_{l,i}&=&w_i(\bm{x}_l)+(\bm{x}_p-\bm{x}_l)\cdot\nabla w_i(\bm{x}_l)
  +\frac{1}{2}(\bm{x}_p-\bm{x}_l)(\bm{x}_p-\bm{x}_l)\colon\nabla\nabla w_i(\bm{x}_l),
  \label{eq:ipvelstep02}
\end{eqnarray}
\begin{eqnarray}
  &&\{\alpha_c(\left<u_i\right>_c-v_{pi})\}(\bm{x}_p)\approx
  \nonumber\\&&\;\;\;\;\;\;\;\;
  \sum_{l=1}^8\left(1-\frac{|x_{p1}-x_{l1}|}{\Delta x}\right)
  \left(1-\frac{|x_{p2}-x_{l2}|}{\Delta x}\right)
  \left(1-\frac{|x_{p3}-x_{l3}|}{\Delta x}\right)W_{l,i}-v_{pi},
  \label{eq:ipvelstep03}
\end{eqnarray}
where $l=1,\cdots,8$ indicates the eight definition points of the velocity around $\bm{x}_p$.
The effect of the second-order derivative is considered in (\ref{eq:ipvelstep02}) and
the linear interpolation (\ref{eq:ipvelstep03}) guarantees the continuity about $\bm{x}_p$.
For the pressure, the linear interpolation
\begin{eqnarray}
  P(\bm{x})\approx
  \sum_{l=1}^8\left(1-\frac{|x_{1}-x_{l1}|}{\Delta x}\right)
  \left(1-\frac{|x_{2}-x_{l2}|}{\Delta x}\right)
  \left(1-\frac{|x_{3}-x_{l3}|}{\Delta x}\right)P_{l}
\end{eqnarray}
is used ($l$ indicates the definition points of the pressure).

To adjust the flow field far away from the particle to the non-averaged field,
the residual stress term $\nabla\cdot\bm{\tau}$ is replaced by $C(\nabla\cdot\bm{\tau})$ where 
$C$ is given as follows:
\begin{eqnarray}
  C(\bm{x})=
   \begin{cases}
     \; \cos\left(\dfrac{\pi}{2}\dfrac{|\bm{x}-\bm{x}_p|}{R+r_d}\right)
     \;\;
         {\rm for~}\; |\bm{x}-\bm{x}_p|\leq R+r_d \\[12pt]
     \; 0 \;\;\;\;\;\;\;\;\;\;\;\;\;\;\;\;\;\;\;\;\;\;\;\;\;\;
     \;\;    {\rm for~}|\bm{x}-\bm{x}_p|>R+r_d
   \end{cases}.
\end{eqnarray}

\subsection{One-way coupling simulation}
In the one-way coupling simulation, the flow field ($\bm{U}_{\rm ud}$, $P_{\rm ud}$)
is theoretically or numerically given and only the particle motion is solved for.
To consider the history effect, the acceleration including the Basset history term is considered.
The equation of motion is given by
\begin{eqnarray}
\frac{d\bm{v}_p}{dt}&=&
\frac{2}{2m_d+m_c}\bigg\{3\pi\rho_c\nu D(\bm{U}_{\rm ud}-\bm{v}_p)
-\frac{\pi D^3}{4}\nabla P_{\rm ud}+(m_d-m_c)\bm{g}
\nonumber\\&&\;\;\;\;\;\;\;\;\;\;\;\;\;\;\;\;\;\;\;\;
+\frac{3}{2}\rho_cD^2\sqrt{\pi\nu} \int_{-\infty}^{t}\frac{1}{\sqrt{t-\tau}}
\frac{d}{d\tau}(\bm{U}_{\rm ud}-\bm{v}_p) d\tau
\bigg\}.
\label{eq:vpcl_l01}
\end{eqnarray}
Eqs.~(\ref{eq:xpcl01}) and (\ref{eq:vpcl_l01}) are solved with
the efficient implicit method proposed by van Hinsberg et al. (2011).
The linear drag force model ($F_{\rm drag}=3\pi\rho_c\nu D|\bm{U}_{\rm ud}-\bm{v}_p|$) is used as they did.
Simulations are also performed without the Basset term for comparison.

To quantify the relevance of the nonlinear drag force model, 
we solve Eqs.~(\ref{eq:vpcl01}) and (\ref{eq:xpcl01}) without the external force, $\bm{h}$, with
a 2nd-order Runge-Kutta method.
Note that the volume force $\bm{h}$ is entirely distributed on the fluid.
The particle Reynolds number used here is given by 
\begin{eqnarray}
{\rm Re}_p=\frac{|\bm{U}_{\rm ud}-\bm{v}_p|D}{\nu}
\label{eq:rep_estimate02}
\end{eqnarray}
instead of Eq.~(\ref{eq:rep_estimate01}); the direction defined by the unit vector 
$\bm{m}$ is parallel to $\bm{U}_{\rm ud}-\bm{v}_p$.

In the present study, to examine the history effect in time when ${\rm Re}_p$ is low,
the following three cases of one-way coupling simulations are performed and 
those are denoted as:
\begin{quote}
\begin{description}
\item[{\rm O-LB:}] including the linear drag, the added mass, the pressure gradient and the Basset terms, see Eq.~(\ref{eq:vpcl_l01}),
\item[{\rm O-L:}\ \ ] same as O-LB but without the Basset term,
\item[{\rm O-NL:}] including the nonlinear drag, the added mass and the pressure gradient, see Eq.~(\ref{eq:vpcl01}).
\end{description}
\end{quote}

\subsection{Traditional two-way coupling simulation}
In the traditional two-way coupling simulation, 
Eq.~(\ref{eq:continuity01}) and the momentum equation
\begin{eqnarray}
\frac{\partial\bm{u}}{\partial t}+\nabla\cdot(\bm{u}\bm{u})=
-\frac{1}{\rho_c}\nabla p + \nu\nabla^2\bm{u}+\bm{h}+\bm{f}_b
\label{eq:NS-tw01}
\end{eqnarray}
are solved without volume-averaging,
where $\bm{f}_b$ is the feedback force from the particle.
Eq.~(\ref{eq:vpcl02}) with $\bm{h}=0$ (as $\bm{h}$ is entirely distributed on the fluid)
and Eq.~(\ref{eq:xpcl01}) are solved for the particle motion.
This traditional two-way coupling simulation is referred to as TT  below.

To distribute the feedback force $\bm{f}_b$, we use the following equation (regularised Dirac delta function)
\begin{eqnarray}\label{eq:spread}
  \bm{f}_b(\bm{x})=
   \begin{cases}
     \; -\dfrac{F_{\rm drag}}{\rho_c}\bm{m}
     K\left\{1+\cos\left(\pi\dfrac{|\bm{x}-\bm{x}_p|}{R+r_d}\right)\right\}
     \;\;
         {\rm for~}\; |\bm{x}-\bm{x}_p|\leq R+r_d \\[12pt]
     \; 0 \;\;\;\;\;\;\;\;\;\;\;\;\;\;\;\;\;\;\;\;\;
     \;\;\;\;\;\;\;\;\;\;\;\;\;\;\;\;\;\;\;\;\;\;\;\;
     \;\;\;\;\;\;\;\;\;\;
     \;\;    {\rm for~}\;|\bm{x}-\bm{x}_p|>R+r_d
   \end{cases}
\end{eqnarray}
where $K$ is the normalisation factor computed as
\begin{eqnarray}
  K=\left[\sum_{ijk} \left\{1+\cos\left(\pi\dfrac{|\bm{x}_{ijk}-\bm{x}_p|}{R+r_d}\right)\right\}
    \Delta x^3\right]^{-1}.
\end{eqnarray}
The direction $\bm{m}$, used in eq.~(\ref{eq:spread}), is determined by $\bm{m}=(\bm{u}(\bm{x}_p)-\bm{v}_p)/|\bm{u}(\bm{x}_p)-\bm{v}_p|$.

The fluid velocity at the particle centre is interpolated with
\begin{eqnarray}
  W_{l,i}&=&u_i(\bm{x}_l)+(\bm{x}_p-\bm{x}_l)\cdot\nabla u_i(\bm{x}_l)
  +\frac{1}{2}(\bm{x}_p-\bm{x}_l)(\bm{x}_p-\bm{x}_l)\colon\nabla\nabla u_i(\bm{x}_l),
  \\
  u_i(\bm{x}_p)&\approx&
  \sum_{l=1}^8\left(1-\frac{|x_{p1}-x_{l1}|}{\Delta x}\right)
  \left(1-\frac{|x_{p2}-x_{l2}|}{\Delta x}\right)
  \left(1-\frac{|x_{p3}-x_{l3}|}{\Delta x}\right)W_{l,i}.
\end{eqnarray}
To compute the drag force by Eq.~(\ref{eq:fdragnl}), the particle Reynolds number is estimated as
\begin{eqnarray}
{\rm Re}_p=\frac{|\bm{u}(\bm{x}_p)-\bm{v}_p|D}{\nu}.
\label{eq:rep_estimate03}
\end{eqnarray}
The numerical procedure with the 2nd-order Runge-Kutta method is employed for both phases.

\subsection{Fully-resolved simulation}
The immersed boundary code originally developed by Breugem (2012) is used for
the fully-resolved simulation.
The continuity equation (\ref{eq:continuity01}) and the following momentum equation
are solved in the whole domain including the regions occupied by the particles:
\begin{eqnarray}
\frac{\partial\bm{u}}{\partial t}+\nabla\cdot(\bm{u}\bm{u})=
-\frac{1}{\rho_c}\nabla p + \nu\nabla^2\bm{u}+\bm{h}+\bm{f}_{\rm IB},
\label{eq:NS-ib01}
\end{eqnarray}
where $\bm{f}_{\rm IB}$ is the body force used to impose 
the no-slip condition on the particle surface.
The particle translational and rotational equations are
\begin{eqnarray}
m_d\frac{d\bm{v}_p}{dt}&=&\oint_S\bm{s}\cdot\bm{n}dS
+\rho_c\int_{V_p}\bm{h}dV+(m_d-m_c)\bm{g},
\label{eq:vpcl-ib01}\\
I_d\frac{d\bm{\Omega}_p}{dt}&=&\oint_S(\bm{x}-\bm{x}_p)\times(\bm{s}\cdot\bm{n})dS
+\rho_c\int_{V_p}(\bm{x}-\bm{x}_p)\times\bm{h}dV,
\label{eq:opcl-ib01}
\end{eqnarray}
where $S$ is the particle surface, 
$\bm{s}=-p\bm{I}+\rho_c\nu(\nabla\bm{u}+\nabla\bm{u}^T)$ the stress tensor.
The force exchange is considered on a set of $N$ Lagrangian points around each particle surface.
The force $\bm{F}_j$ at the $j$th Lagrangian point is distributed on the fluid as
\begin{eqnarray}
\bm{f}_{\rm IB}(\bm{x})=\sum_{j=1}^{N}\bm{F}_j\delta_d(\bm{x}-\bm{x}_j)\Delta V_j,
\label{eq:ibf01}
\end{eqnarray}
where $\delta_d$ is a regularised Dirac delta function and 
$\Delta V_j$ the volume of the Lagrangian grid cell.
In the simulation, Eqs.~(\ref{eq:vpcl-ib01}) and (\ref{eq:opcl-ib01}) are converted to
\begin{eqnarray}
m_d\frac{d\bm{v}_p}{dt}&=&-\rho_c\sum_{j=1}^{N}\bm{F}_j\Delta V_j
+\rho_c\frac{d}{dt}\left(\int_{V_p}\bm{u}dV\right)+(m_d-m_c)\bm{g},
\label{eq:vpcl-ib02}\\
I_d\frac{d\bm{\Omega}_p}{dt}&=&-\rho_c\sum_{j=1}^{N}(\bm{x}_j-\bm{x}_p)\times\bm{F}_j\Delta V_j
+\rho_c\frac{d}{dt}\left(\int_{V_p}(\bm{x}_l-\bm{x}_p)\times\bm{u}dV\right).
\label{eq:opcl-ib02}
\end{eqnarray}
The three-step Runge-Kutta method is used for the time integration. 
More details can be found in Breugem (2012) and Lambert et al. (2013).

\section{Numerical results}
In the VA simulation, the history and lift forces are captured without any specific treatment 
as this method intrinsically incorporates the effect of the flow perturbation through the volume-averaging. 
In Sec~4.1, the particle settling problem in a stationary fluid is simulated to show how the history force is represented. 
The lift force on a rotating particle is instead the focus of Sec 4.2.
For the study of the turbulence modulation by particles of comparable size to the Kolmogorov length scale, 
the interaction between the particle and a vortex element should be precisely represented. 
In Sec~4.3, therefore, the applicability of the VA simulation for the Taylor-Green vortex is investigated. 
As the fundamental validation for different density ratios, the particle motion in the smallest periodic unit is simulated without gravity. 
Finally, in Sec~4.4, to highlight the importance of the two-way coupling simulation for a vortical flow, 
we show the particle trajectory and the inter-particle interaction in an array of Taylor-Green units with gravity.

Throughout this study, grid resolutions of 
$D/\Delta x=24$ for the fully-resolved simulations
and $D/\Delta x=2$ for the VA simulations are commonly employed.
Note that the number of grid points is $12^3$ times lower for 
the VA simulation with respect to the fully-resolved simulation
and consequently the time step $\Delta t$ is $20$ times larger.
Therefore, the total computational cost is $\mathcal{O}(10^4)$ times lower with the VA model.
The computational domain is rectangular of lengths $l_1,l_2$ and $l_3$ in the $x_1,x_2$ and $x_3$-directions.
Periodic boundary conditions are applied in all the directions.
The motion of the particle is confined in the $x_1$-$x_2$ plane due to 
the symmetry of the flows studied.
The choice of $R=0.75D$ and $D/\Delta x=2$ is reasonable according to our previous study (Fulada et al., 2016).
The effects of these parameters are further discussed in Appendix~E.

\subsection{History effect on the settling particle}\label{sec:result_settling1}
A single particle settling in a stationary fluid is studied by one-way coupling simulations, 
the VA approach and the IBM simulation.
The fluid and particle velocities are initially set to $0$.
Gravity acts in the negative $x_2$-direction and the external force, $\bm{h}$, is neglected. 
The importance of gravity is characterised by the Galileo number defined as
\begin{eqnarray}
{\rm Ga}=\frac{\sqrt{\left(\dfrac{\rho_d}{\rho_c}-1\right)|\bm{g}|D^3}}{\nu}.
\label{eq:Galileo01}
\end{eqnarray}
In the following, using Eq.~(\ref{eq:fdragnl}), ${\rm Ga}$ is related to the particle Reynolds number 
based on the particle terminal velocity as
\begin{eqnarray}
{\rm Ga}=\sqrt{18{\rm Re}_{\rm term}(1+0.15{\rm Re}_{\rm term}^{0.687})}.
\label{eq:Galileo02}
\end{eqnarray}

The following set of parameters are used for the simulations:
$(l_1/D,l_2/D,l_3/D)=(16,32,16)$, $\rho_d/\rho_c=100$ and ${\rm Ga}=8.44$ (corresponding to ${\rm Re}_{\rm term}=3$).
The time step is $(\nu/D^2)\Delta t=1.19\times 10^{-4}$ for the fully-resolved simulation
and $(\nu/D^2)\Delta t=2.39\times 10^{-3}$ for the other cases.
The number of grid points is $384\times 768\times 384$ for the fully-resolved simulation
and $32\times 64\times 32$ for the VA simulation.

\begin{figure}[!h]
\centering
 \psfrag{t}[s][s][1]{$t\nu/D^2$}
 \psfrag{v}[c][c][1]{$v_{p2}D/\nu$}
 \psfrag{model}[r][r][1]{present model}
 \psfrag{oneway-NL}[r][r][1]{one-way with Eq.~(\ref{eq:vpcl01})}
 \psfrag{oneway-L}[r][r][1]{one-way with Eq.~(\ref{eq:vpcl_l01})}
 \psfrag{oneway-LB}[r][r][1]{one-way with Eq.~(\ref{eq:vpcl_l01})}
\begin{overpic}[scale=0.6, angle=-90]{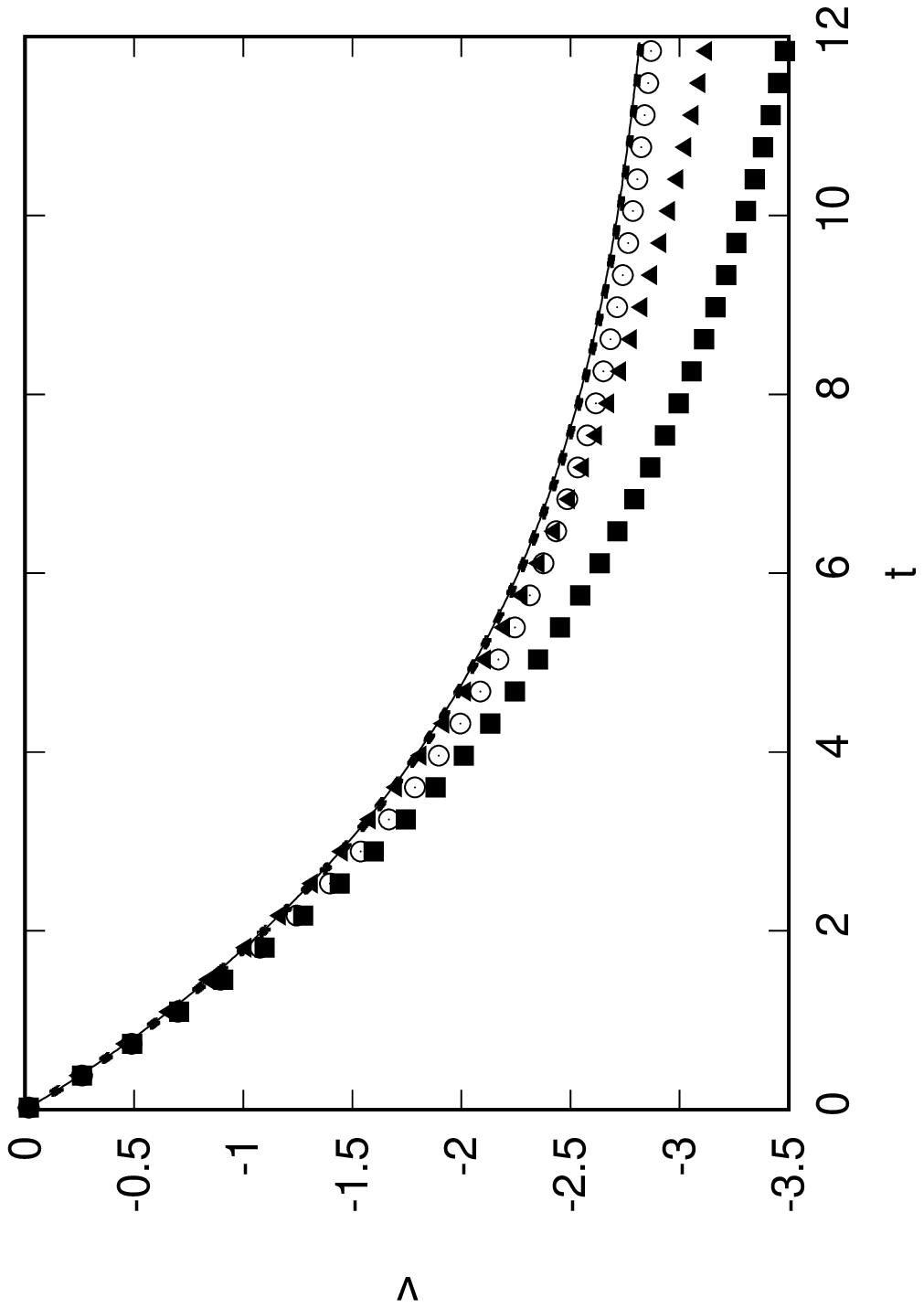}
  \put(0,65){(a)}
\end{overpic}\\
\begin{overpic}[scale=0.6, angle=-90]{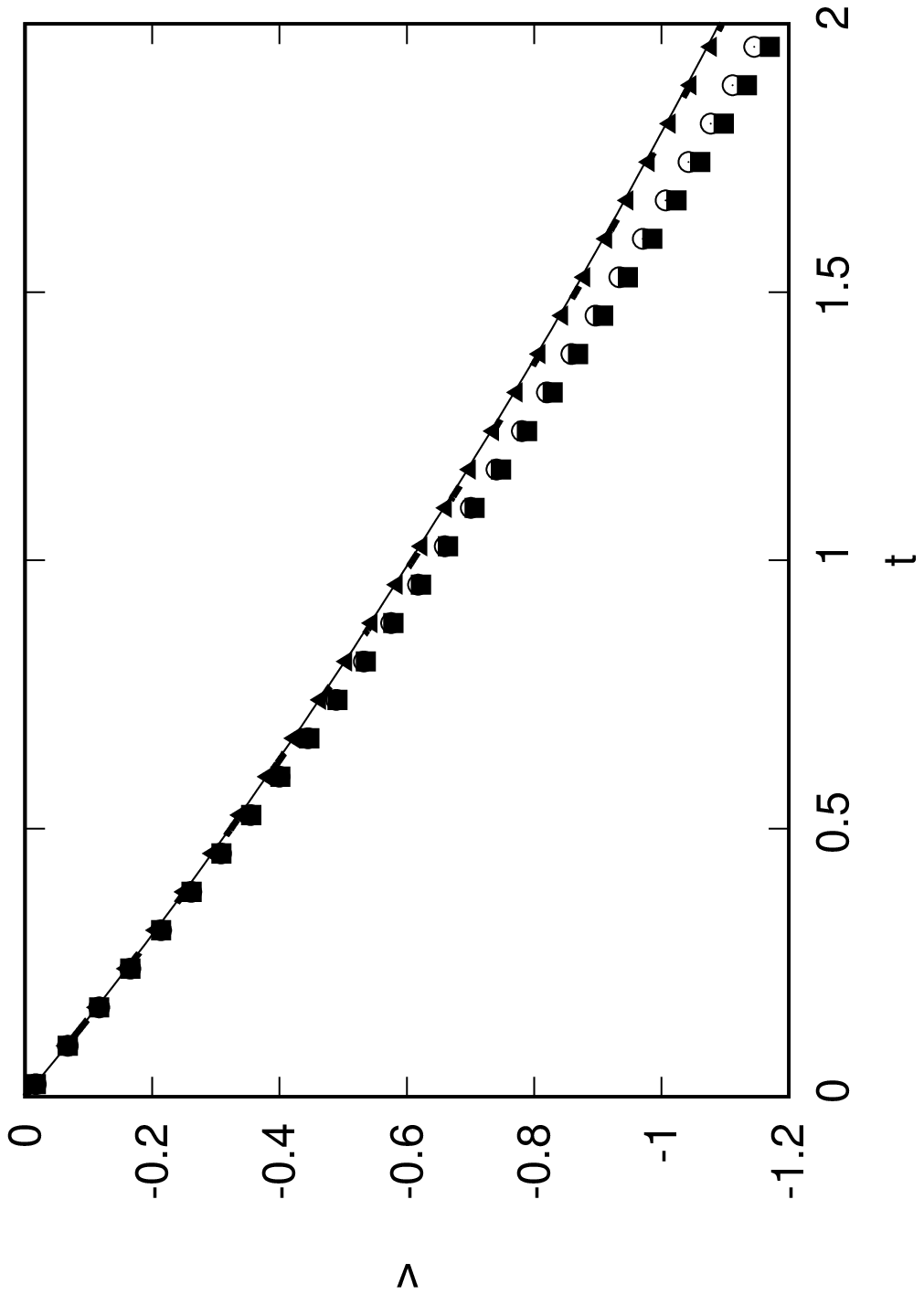}
  \put(0,65){(b)}
\end{overpic}
\caption{(a) Time evolution of the particle settling velocity in a fluid at rest.
Solid line, VA simulation; dashed line, fully-resolved simulation;
circle, O-NL simulation; 
filled triangle, O-LB simulation; 
filled square, O-L simulation.
(b) Enlarged view of the same data at the early stage.}
\label{fig:falling_vel}
\end{figure}

Figure~\ref{fig:falling_vel} shows the time evolution of the dimensionless 
particle settling velocity $v_{p2}D/\nu$.
First we note that the solid and dashed lines almost overlap with each other and
the proposed VA and the fully-resolved simulations show good agreement.
Around $(\nu/D^2)t=12$, the result of the O-NL simulation, including drag, added-mass and pressure gradient, shows
better agreement with that of the fully-resolved simulation due to the nonlinear drag model (\ref{eq:fdragnl}).
On the other hand, the O-LB simulation, including the history effect, 
shows better agreement with the fully-resolved case only for the earlier stage.
Therefore, the history force is essential to correctly model the initial transient stage, which is captured in the VA simulation.
In Fig.~\ref{fig:falling_vel}(b), focusing on the initial stages of the particle motions,
the difference between the results of the two one-way coupling simulations without the Basset term (O-L and O-NL)
is small because the nonlinear effect in the drag force is not significant
at the initial stage when the particle Reynolds number is low.

As the boundary layer thickness at the beginning of the settling  
is smaller than that in the steady flow,
the friction drag of the unsteady flow becomes larger in the developing stage.
In the VA simulation, smaller boundary layer thickness corresponds to 
larger $|\alpha_c(\left<\bm{u}\right>_c-\bm{v}_p)|$ and
the history effect is qualitatively reflected in the drag force.
This also explains why the result of the VA simulation shows quantitatively 
good agreement with the fully-resolved simulation.

\subsection{Lift force induced by particle rotation}\label{sec:result_magnus1}
The VA simulation of the flow around a rotating particle is carried out to test 
the capability of capturing the transversal forces.
The uniform velocity $\bm{u}=(U_{\rm init},0,0)$ is given as the initial condition for the fluid flow.
The particle centre is fixed in space and the angular velocity is kept constant 
to $\bm{\Omega}_p=(0,\Omega_{\rm const},0)$,
thus the particle motion, Eqs.~(\ref{eq:vpcl01}), (\ref{eq:xpcl01}) 
and (\ref{eq:opcl01}), does not need to be solved.
The Reynolds numbers and the angular velocities are varied in the following range:
$U_{\rm init}D/\nu=1,5,10,20$ and 
$\Omega_{\rm const}D/U_{\rm init}=0.196,0.393$, with
gravity and the external forces set to zero.
The size of the computational domain is $(l_1/D,l_2/D,l_3/D)=(64,32,16)$ and
the number of grid points is $128\times 64\times 32$.
The time step is $(U_{\rm init}/D)\Delta t=2.81\times 10^{-2}$.
The wake of the particle reaches the particle position
around $(U_{\rm init}/D)t=64$ due to the periodic boundary condition.
The force is thus examined at $(U_{\rm init}/D)t=28.1$ so that effects from the re-entering wake are avoided.

\begin{figure}[!h]
\centering
 \psfrag{cdcl}[s][s][1]{$C_D$, $C_L$}
 \psfrag{rep}[c][c][1]{${\rm Re}_p$}
\begin{overpic}[scale=0.77, angle=-90]{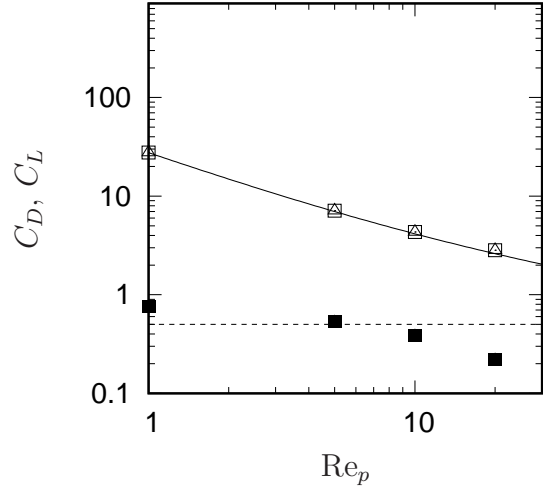}
\end{overpic}
\caption{Drag and lift coefficients on a rotating particle for different Reynolds numbers by the VA simulations, 
indicated by open and filled symbols.
Square symbols, rotation rate $\Omega_{\rm const}D/U_{\rm init}=0.196$; triangle symbols, $\Omega_{\rm const}D/U_{\rm init}=0.393$.
The solid line shows the drag coefficient obtained from Eq.~(\ref{eq:fdragnl})
and the dashed line represents $C_L=0.5$.
Square and triangle symbols almost overlap with each other.}
\label{fig:magnus_cdcllist}
\end{figure}

Based on the components of the estimated fluid force $\bm{F}$,
the drag and lift coefficients, $C_D$ and $C_L$, are defined as follows: 
\begin{eqnarray}
C_D=\frac{F_1}{\dfrac{\pi}{8}\rho_cU_{\rm init}^2D^2},\;\;\;
C_L=\frac{-F_2}{\dfrac{\pi}{8}\rho_cU_{\rm init}\Omega_{\rm const}D^3}.
\label{eq:cdcldef}
\end{eqnarray}
The drag and lift coefficients obtained in the VA simulation are plotted in Fig.~\ref{fig:magnus_cdcllist}.
Note that the magnitude of the angular velocity does not influence the two coefficients.
The drag coefficient estimated in the present simulation shows good agreement with that based 
on Eq.~(\ref{eq:intf_unif01}) (solid line).
Therefore, the effect of the rotation on the drag force is small as supported by previous researches
(Sridhar and Katz, 1995; Bagchi and Balachandar, 2002).
As for the lift force, the signs of $\alpha_c\left<u_2\right>_c$ at the particle centre and $F_2$ should be the same
according to the result that the contribution of the friction lift is in the same direction as $F_2$ (Kurose and Komori, 1999).
Therefore, the present force estimation (Sec.~2.3.1), the directions of the drag force 
and $\alpha_c(\left<\bm{u}\right>_c-\bm{v}_p)$ being the same,
is capable of capturing the direction of the lift force.
In the VA simulation,
the interaction force model for the particle rotation, Eq.~(\ref{eq:intf_rot01}), 
induces $\alpha_c\left<u_2\right>_c <0$ and thus the lift force $F_2<0$.
According to the theoretical study by Rubinow and Keller (1961),
the lift coefficient is $C_L=1$ for ${\rm Re}_p\ll 1$.
On the other hand, numerical studies at ${\rm Re}_p=\mathcal{O}(10)$ have shown an estimate of $C_L\approx 0.5$ 
(Bagchi and Balachandar, 2002; Bluemink et al., 2010).
The VA simulation captures the direction of the lift force generated by the particle rotation,
and the magnitude quantitatively agrees with the previous results for ${\rm Re}_p < 10$
under the setting of this study.

\subsection{Vortical flow without gravity for three density ratios}\label{sec:result_vor_nograv1}
To study the interaction between a particle and a vortex,
the Taylor-Green vortex is used as the background undisturbed flow.
The smallest unit structure of the Taylor-Green vortex is considered
to compare the results of relatively simple particle motions from different simulations.
The directions of the Cartesian coordinates ($x_1,x_2,x_3$) are determined 
so that the velocity components of the undisturbed flow are
\begin{eqnarray}
\bm{U}_{\rm ud}\cdot\bm{e}_1=A\sin\left(\frac{x_2}{L}\right),\;\;\;
\bm{U}_{\rm ud}\cdot\bm{e}_2=-A\sin\left(\frac{x_1}{L}\right),
\label{eq:flowvel_TG1}
\end{eqnarray}
where the velocity $A$ and the length $L$ define the vortex intensity and size.
The period in the $x_1$ and $x_2$-directions is $2\pi L$ and the Reynolds number ${\rm Re}=AL/\nu =18$.
According to Jim\'{e}netz et al. (1993), the intensity of a typical vortex in isotropic turbulence 
is correlated as $\Gamma/\nu\approx 18\sqrt{{\rm Re}_{\lambda}}$,
where $\Gamma$ is the circulation of the vortex and ${\rm Re}_{\lambda}$ the Reynolds number based on the Taylor length scale.
The present case (${\rm Re}=18$), where the circulation of one vortex is $\Gamma=16AL$, 
corresponds therefore to  ${\rm Re}_{\lambda}\approx 250$.
The size of the computational domain is $l_1/L=l_2/L=l_3/L=2\pi$ and the particle diameter $D/L=2\pi/16$.
As the Kolmogorov length scale $\eta$ is around eight times smaller than the diameter of the most intense vortices in turbulence
(Jim\'{e}netz et al., 1993),
the present particle diameter is considered as a model of the case $D\sim\eta$.
The flow is maintained by the external force
\begin{eqnarray}
\bm{h}\cdot\bm{e}_1= \frac{A\nu}{L^2}\sin\left(\frac{x_2}{L}\right),\;\;\;
\bm{h}\cdot\bm{e}_2=-\frac{A\nu}{L^2}\sin\left(\frac{x_1}{L}\right).
\label{eq:flowext_TG1}
\end{eqnarray}
The number of grid points is $384\times 384\times 384$ for the fully-resolved simulation
and $32\times 32\times 32$ for the VA and the TT simulations.
The time step is $(A/L)\Delta t=3.31\times 10^{-4}$ for the fully-resolved simulation
and $(A/L)\Delta t=6.63\times 10^{-3}$ for the other two methods.
Three different density ratios ($\rho_d/\rho_c=1,10,1000$) and two different initial particle positions are examined in the following.
The corresponding Stokes numbers ${\rm St}=\rho_dD^2A/(18\rho_c\nu L)$ are 0.154, 1.54 and 154.
The initial velocity of the particle is set to be the same as the undisturbed fluid velocity at the particle centre
and the initial angular velocity is 0.

\subsubsection{$\rho_d/\rho_c=1000$}\label{sec:result_vor_nograv1_1000}
When the initial particle position is $(x_1/L,x_2/L,x_3/L)=(\pi/2,\pi/2,0)$,
the particle trajectory follows the straight line defined by $dx_2/dx_1=-1$
through the periodic boundaries and the particle does not rotate. 
To highlight the difference between the VA simulation and the TT simulation,
the  VA simulation is repeated without considering the pressure gradient, particle rotation and 
the external force on the particle.
This simplified VA simulation is referred to as SVA simulation.

\begin{figure}[!h]
\centering
 \psfrag{t}[s][s][1]{$(A/L)t$}
 \psfrag{vx}[c][c][1]{$v_{p1}/A$}
\begin{overpic}[scale=0.77, angle=-90]{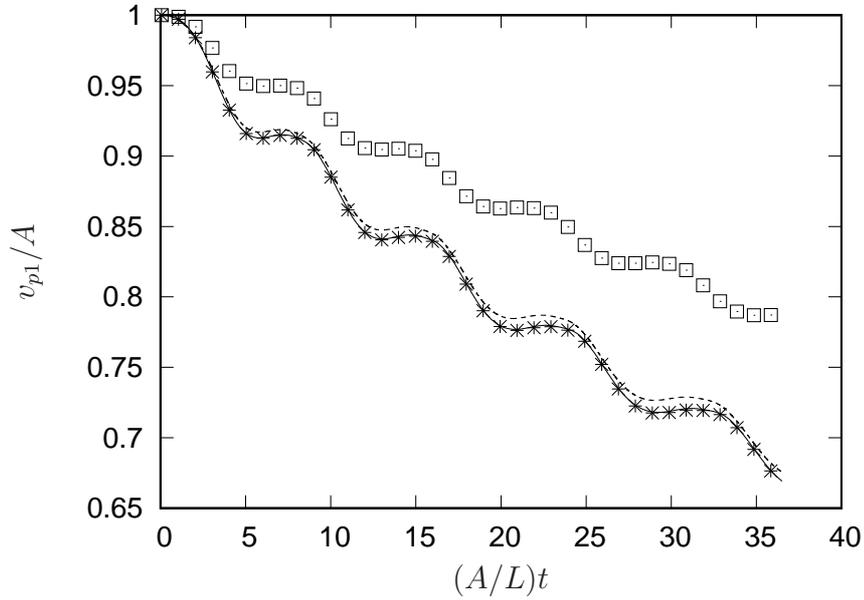}
\end{overpic}
\caption{Time evolution of the particle velocity $v_{p1}$ in a vortical flow for $\rho_d/\rho_c=1000$.
Solid line, VA simulation; 
asterisk, SVA simulation;
dashed line, fully-resolved simulation;
square, TT simulation}
\label{fig:job_init1_den1000_vx-t}
\end{figure}

The time evolution of the particle velocity $v_{p1}$ from these different simulations 
are compared in Fig.~\ref{fig:job_init1_den1000_vx-t}.
The results of the VA and SVA simulations are similar to that of the fully-resolved IBM simulation, which we take as the reference case.
On the other hand, the particle behaviour predicted by the TT simulation exhibits large difference from the reference case.
One of the most significant differences between the SVA and the TT simulations is
the estimation of the drag force. 
As the flow disturbance is non-negligible for the finite-size particle,
the force estimation according to Eq.~(\ref{eq:rep_estimate03}) 
without considering the local flow disturbance underestimates the drag force 
and results in the smaller acceleration of the particle in the TT simulation.

\begin{figure}[!h]
\centering
 \psfrag{x}[s][s][1]{$x_1/\pi L$}
 \psfrag{y}[c][c][1]{$x_2/\pi L$}
\begin{overpic}[scale=0.5, angle=-90]{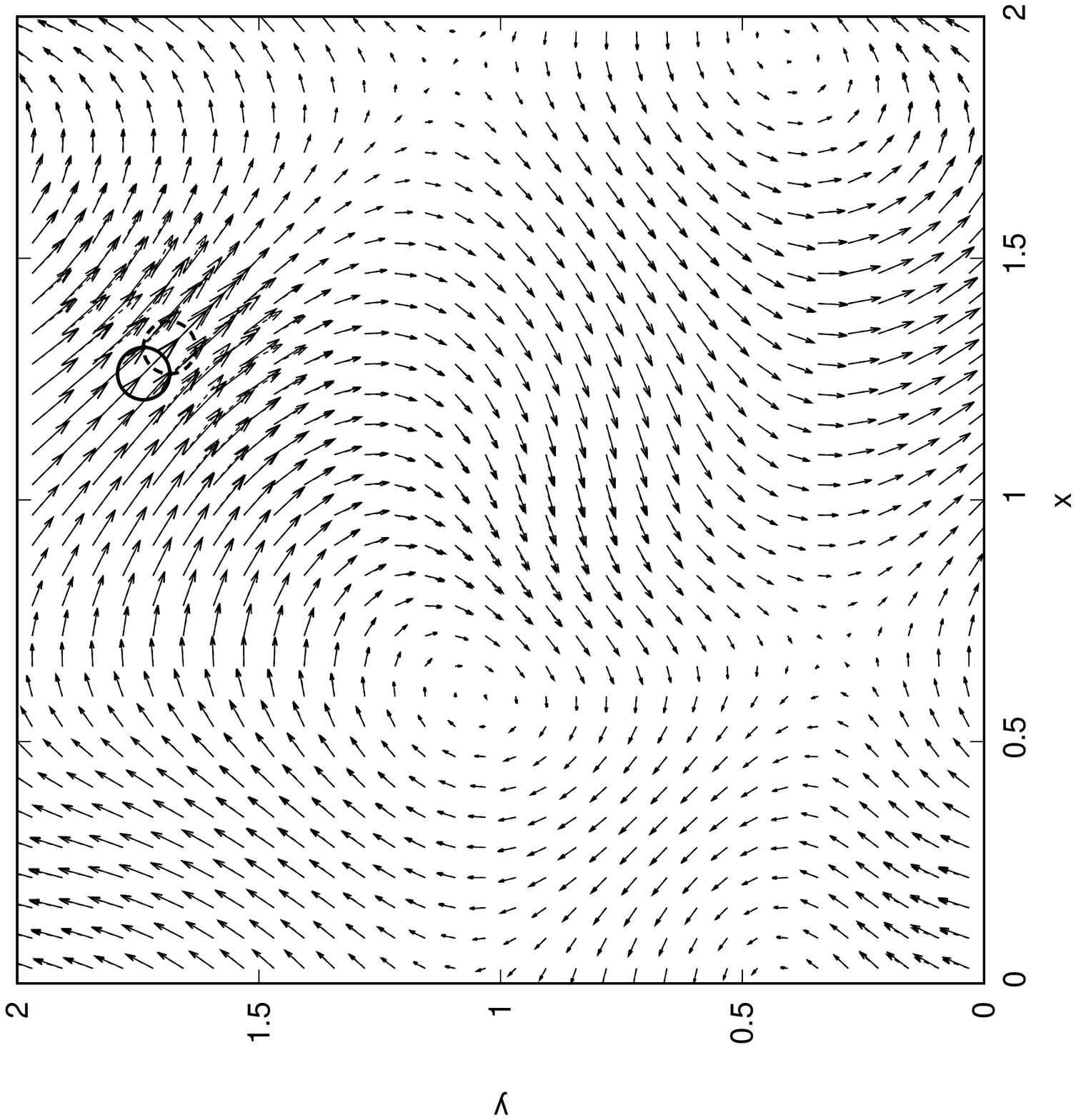}
  \put(10,65){$(a)$}
\end{overpic}\\
\begin{overpic}[scale=0.5, angle=-90]{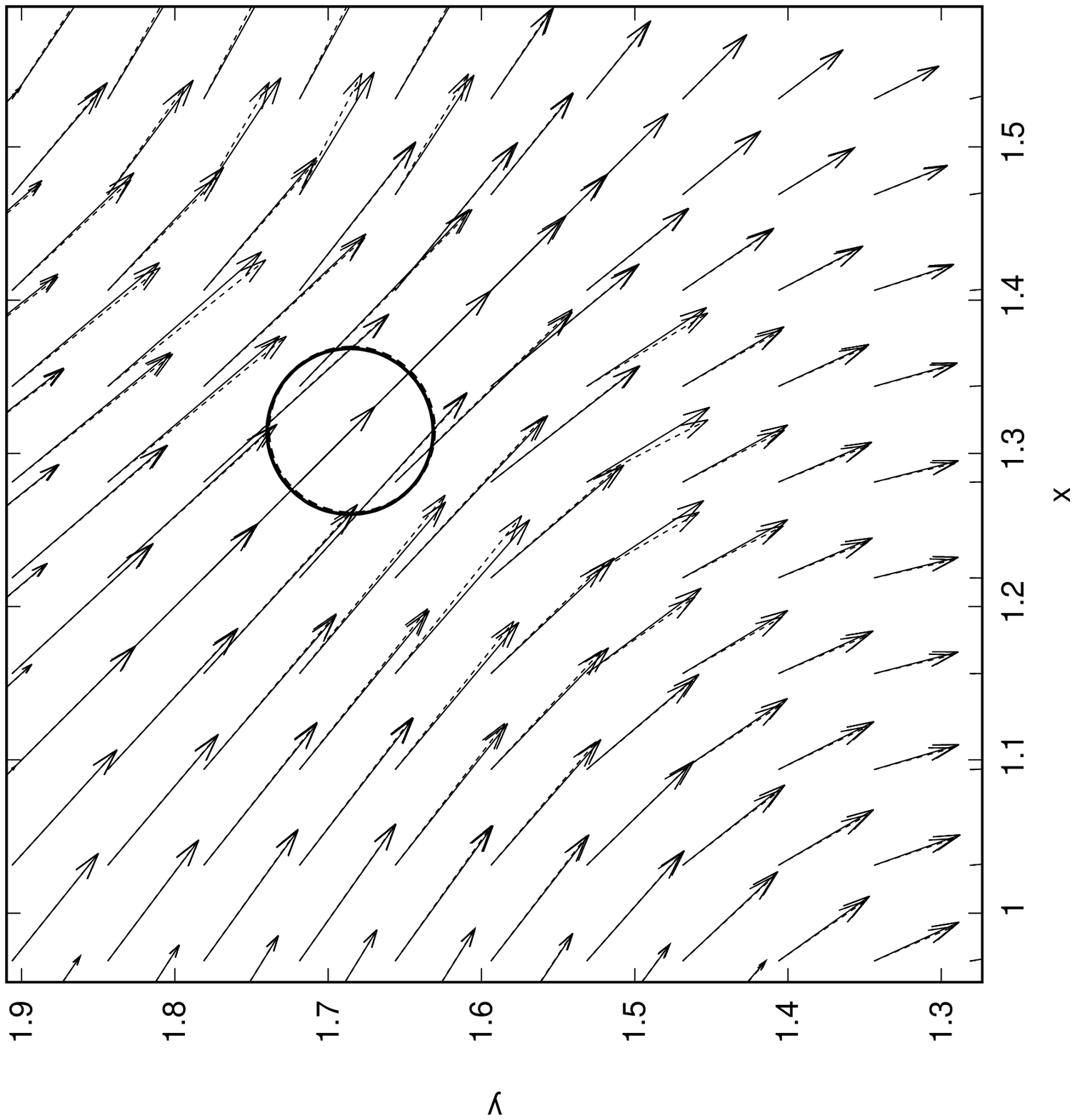}
  \put(10,65){$(b)$}
\end{overpic}
\caption{(a) Disturbance velocity field induced by a particle transported in a Taylor-Green vortex in the $x_1$-$x_2$ 
cross-section cutting through the particle at $(A/L)t=33.13$ for $\rho_d/\rho_c=1000$.
Solid and dashed vectors represent the results of the VA simulation and the fully-resolved simulation.
The circles show the positions of the corresponding particles (by the VA and the fully-resolved simulations).
(b) Enlarged view of the disturbance velocity field around the particle when the time of the VA simulation is changed to $(A/L)t=33.36$
to adjust the particle position to that of the fully-resolved simulation.}
\label{fig:job_init1_den1000_vec}
\end{figure}

To investigate the effect of the particle on the vortex,
we define the induced flow disturbance as $(\alpha_c\left<\bm{u}\right>_c+\alpha_d\left<\bm{v}\right>_d-\bm{U}_{\rm ud})$.
For comparison, the induced flow disturbance for the fully-resolved simulation is defined as
$(\alpha_c\left<\bm{u}\right>_c+\alpha_d\left<\bm{v}\right>_d-\bm{U}_{\rm ud})$ using the local velocities
only in the region where $\alpha_c>0$,
while $(\bm{u}-\bm{U}_{\rm ud})$ is used in the other region.
Figure~\ref{fig:job_init1_den1000_vec}(a) shows the induced flow disturbances at time $(A/L)t=33.13$ in the $x_1$-$x_2$ 
cross-section cutting through the particle 
for both the VA simulation (solid arrow) and the fully-resolved simulation (dashed arrow).
This figure indicates that the disturbances at larger scales than the particle size are very close to each other.
Relatively larger differences are found in the area closer to the particle due to the difference in the position of the particle.
As shown in Fig.~\ref{fig:job_init1_den1000_vec}(b),
by extracting the data from the VA simulation at time $(A/L)t=33.36$, to match the particle position to that of the fully resolved case,
the difference in the flow disturbance becomes smaller.
To summarise, the VA simulation shows a better agreement with the fully-resolved results 
for both the flow disturbance and the particle motion in comparison to the one-way coupling and the TT models.

\subsubsection{$\rho_d/\rho_c=10$}\label{sec:result_vor_nograv1_10}
\begin{figure}[!h]
\centering
 \psfrag{x}[s][s][1]{$x_1/\pi L$}
 \psfrag{y}[c][c][1]{$x_2/\pi L$}
\vspace{-10mm}
\begin{overpic}[scale=0.82, angle=-90]{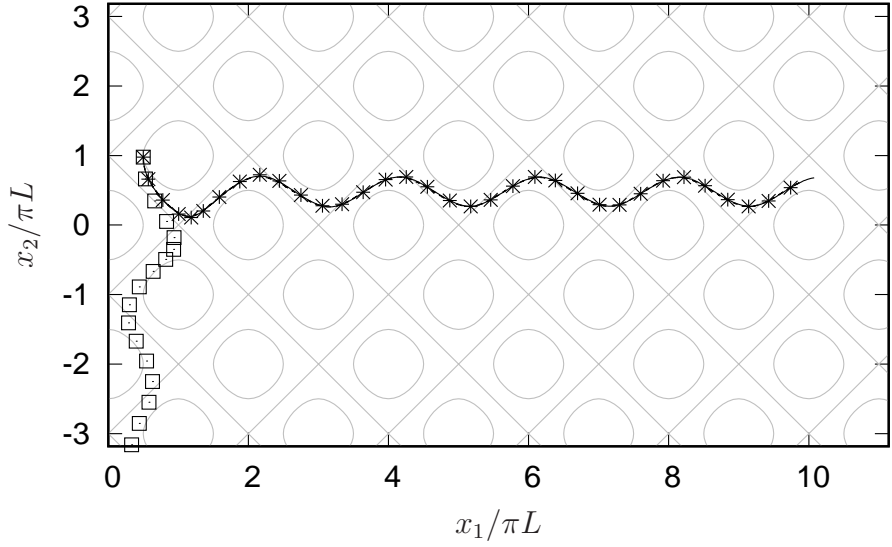}
\end{overpic}
\vspace{-10mm}
\caption{Trajectory of a particle with density ratio $\rho_d/\rho_c=10$ in an array of Taylor-Green vortices.
Solid line, VA simulation; 
asterisk, SVA simulation;
dashed line, fully-resolved simulation;
square, TT simulation;
The grey lines show the streamlines of the undisturbed flow.}
\label{fig:job_init2_den0010_xy}
\end{figure}

The initial particle position is given as $(x_1/L,x_2/L,x_3/L)=(\pi/2,\pi,0)$ so that 
the particle trajectory bends due to the vortical flow.
The particle trajectories for the different simulation methods are compared in Fig.~\ref{fig:job_init2_den0010_xy}.
The result of the VA simulation is very similar to that of the reference fully-resolved IBM simulation.
The effects of the particle rotation, pressure gradient and external force are not significant 
as the result of the SVA simulation is also very close to the two previous cases.
As discussed in Sec.~\ref{sec:result_vor_nograv1_1000},
the drag force estimated in the TT simulation is 
smaller than that in the VA simulation, which gives smaller acceleration in the $x_1$-direction at the early stage.
In the present case where the effect of the pressure gradient is not so significant,
the VA simulation effectively reproduces the curved particle trajectory with significantly less spatial resolution.

\begin{figure}[!h]
\centering
 \psfrag{t}[s][s][1]{$(A/L)t$}
 \psfrag{omega}[c][c][1]{$\Omega_{p3}L/A$}
\vspace{-10mm}
\begin{overpic}[scale=0.77, angle=-90]{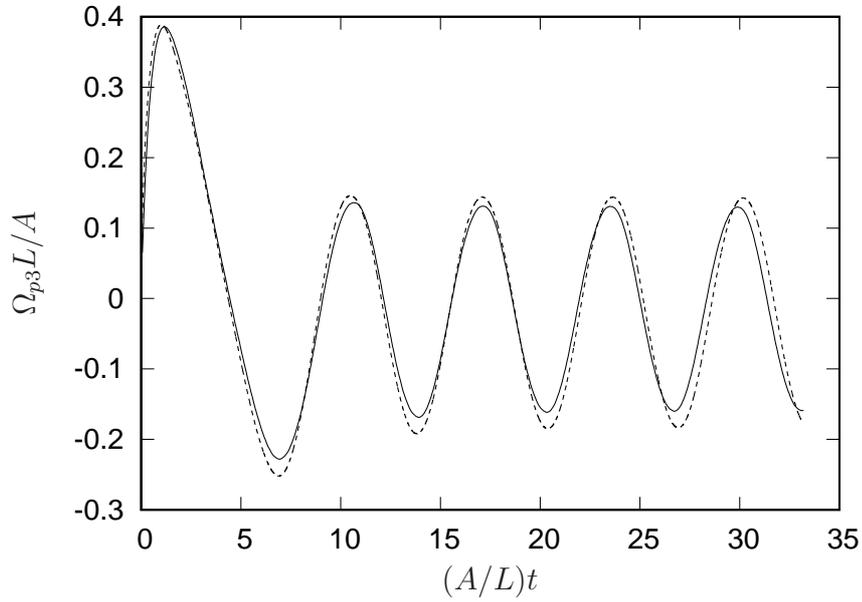}
\end{overpic}
\caption{Time evolution of the angular velocity $\Omega_{p3}$ for a particle with density ratio $\rho_d/\rho_c=10$ transported in an array of Taylor-Green vortices.
Solid line, VA simulation; 
dashed line, fully-resolved simulation.}
\label{fig:job_init2_den0010_oz-t}
\end{figure}

Figure~\ref{fig:job_init2_den0010_oz-t} shows the time evolution of the angular velocity $\Omega_{p3}$.
The result of the VA simulation shows good agreement with that 
of the fully-resolved simulation.
Therefore, the contribution of the vorticity to $\Omega_{p3}$ is reasonably reproduced by the proposed model.

\begin{figure}[!h]
\centering
 \psfrag{x}[s][s][1]{$x_1/\pi L$}
 \psfrag{y}[c][c][1]{$x_2/\pi L$}
\begin{overpic}[scale=0.5, angle=-90]{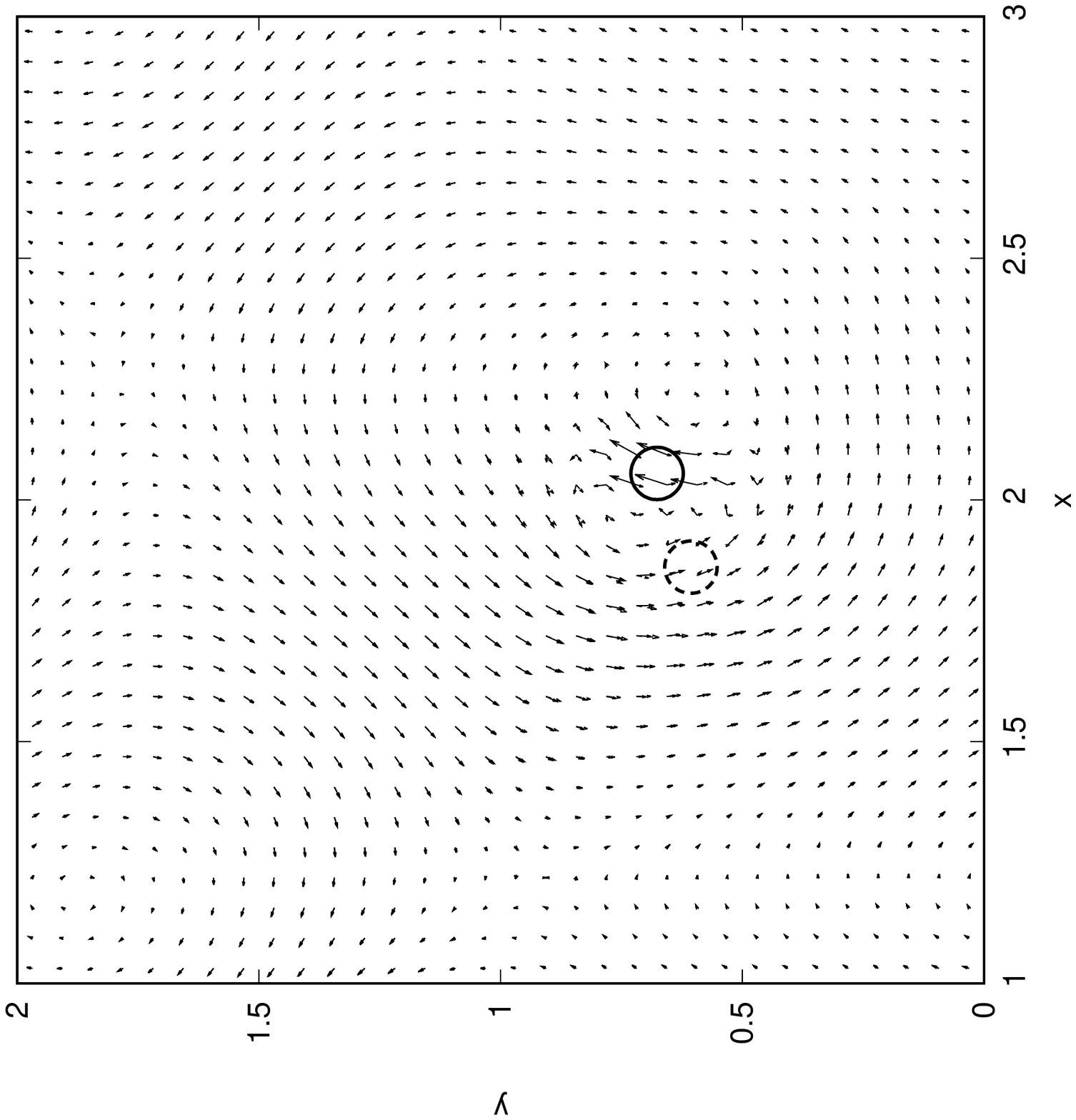}
  \put(10,65){$(a)$}
\end{overpic}\\
\begin{overpic}[scale=0.5, angle=-90]{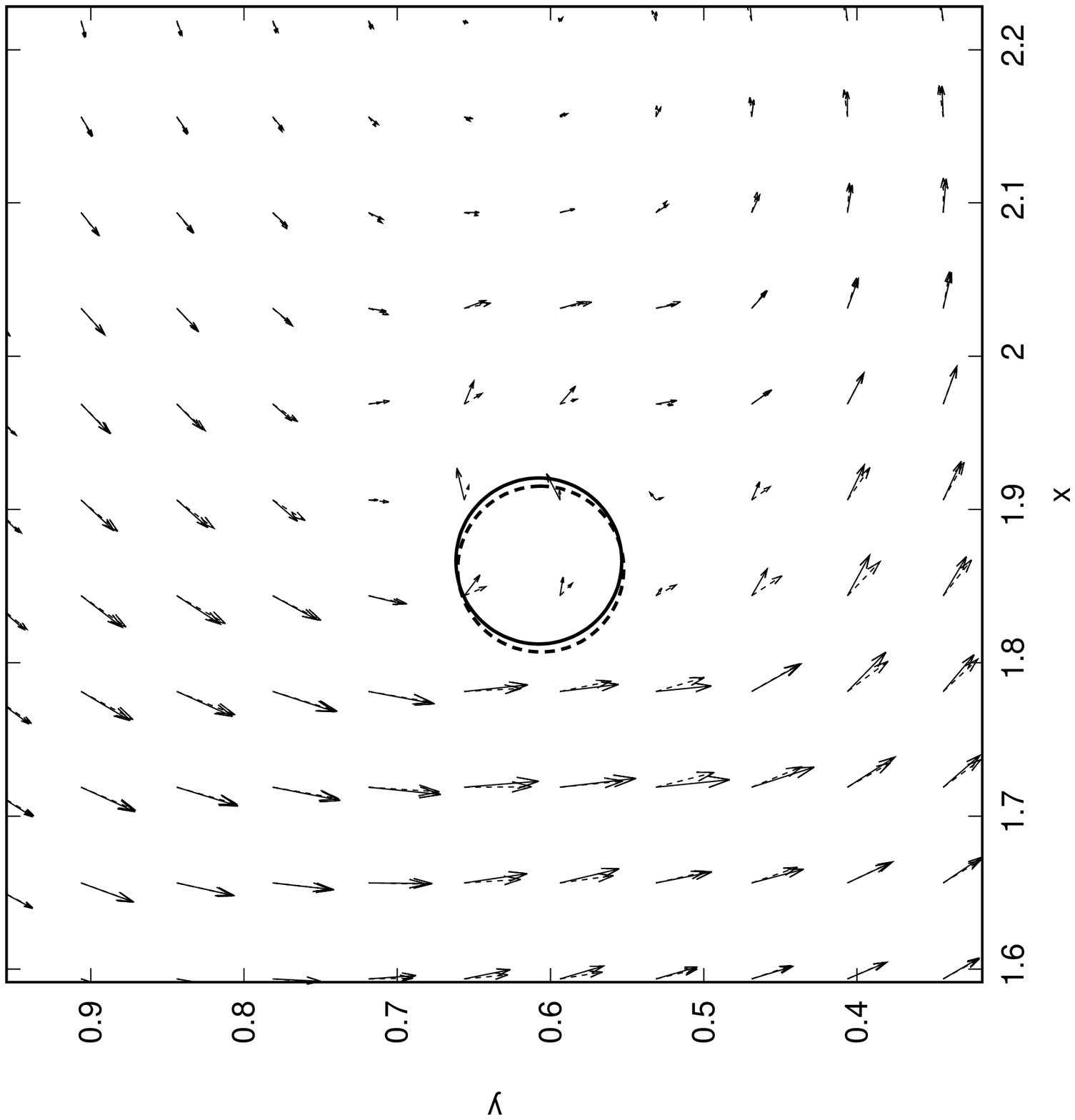}
  \put(10,65){$(b)$}
\end{overpic}
\caption{(a) Disturbance velocity field in the $x_1$-$x_2$ cross-section cutting through the particle 
at $(A/L)t=33.13$ for a particle with $\rho_d/\rho_c=10$ transported in a Taylor-Green array of vortices.
Solid and dashed vectors represent the results of VA simulation and the fully-resolved IBM simulation.
The circles show the positions of the corresponding particles (from the VA and the fully-resolved simulations).
(b) Enlarged view of the disturbance velocity field around the particle when  
the time for VA simulation is changed to $(A/L)t=32.53$ 
to adjust the particle position to that of the fully-resolved simulation.}
\label{fig:job_init2_den0010_vec}
\end{figure}

Finally, Fig.~\ref{fig:job_init2_den0010_vec} (a) shows the induced disturbance velocity field at $(A/L)t=33.13$
in the $x_1$-$x_2$ cross-section cutting through the particle for the VA and the fully-resolved IBM simulations.
The disturbances at larger scales than the particle size show good agreement with each other.
As shown in Fig.~\ref{fig:job_init2_den0010_vec} (b),
the disturbances around the particles give an even better agreement when 
the particle position of the VA simulation is
adjusted to that of the fully-resolved simulation by slightly changing the time ($(A/L)t=32.53$).

\subsubsection{$\rho_d/\rho_c=1$}\label{sec:result_vor_nograv1_1}
\begin{figure}[!h]
\centering
 \psfrag{x}[s][s][1]{$x_1/\pi L$}
 \psfrag{y}[c][c][1]{$x_2/\pi L$}
\begin{overpic}[scale=0.77, angle=-90]{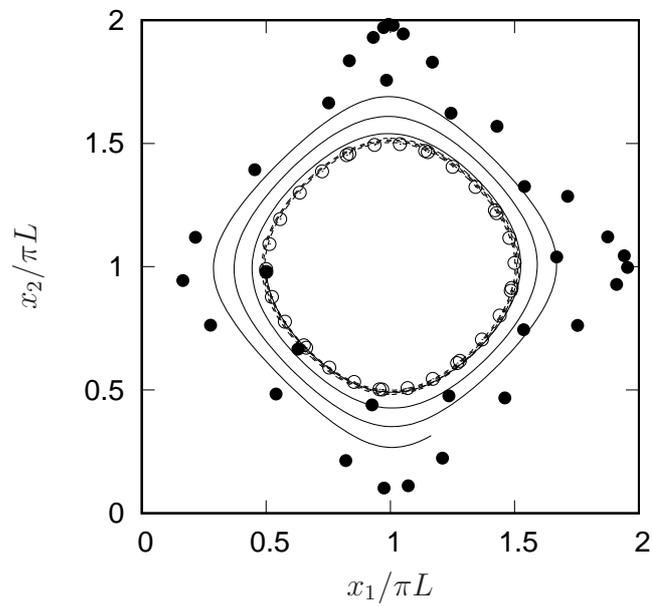}
\end{overpic}
\caption{Trajectory of a particle of density equal to that of the fluid, 
$\rho_d/\rho_c=1$, in an array of Taylor-Green vortices.
Solid line, VA simulation; 
dashed line, fully-resolved IBM simulation;
open circle, O-NL simulation;
filled circle, O-NL simulation further neglecting pressure gradient and added-mass forces.}
\label{fig:job_init2_den0001_xy}
\end{figure}

\sloppy
We next shortly consider particles of density equal to that of the fluid 
with initial particle position $(x_1/L,x_2/L,x_3/L)=(\pi/2,\pi,0)$.
The particle trajectories are compared in Fig.~\ref{fig:job_init2_den0001_xy}.
As the density ratio is 1, the particle velocity fluctuations are relatively large.
The result of the O-NL simulation including all the forces except for the
history effect and external force, Eq.~(\ref{eq:vpcl01}),
shows good agreement with that of the IBM simulation; the streamlines are almost closed.
On the other hand, the O-NL simulation further neglecting the pressure gradient and added mass forces, Eq.~(\ref{eq:vpcl02}),
shows a totally different trend, suggesting that the pressure gradient gives an important contribution.
The result of the VA simulation is also different from that of the fully-resolved IBM simulation.
Therefore, the estimation of the fluid force needs to be improved for the case
where the pressure force is dominant and the particle velocity fluctuations are large.

\subsection{Effects of two-way coupling in vortical flow with gravity}\label{sec:result_vor_grav1}
The settling motion of a particle is investigated in an array of the Taylor-Green units.
The flow configuration is
\begin{eqnarray}
U_{{\rm ud}1}= A\sin\left(\frac{x_1}{L}\right)\cos\left(\frac{x_2}{L}\right),\;\;\;
U_{{\rm ud}2}=-A\cos\left(\frac{x_1}{L}\right)\sin\left(\frac{x_2}{L}\right),
\label{eq:flowvel_TG2}
\end{eqnarray}
as often used (see e.g.\ Maxey, 1987; Bergougnoux et al., 2014).
Note that $A$ and $L$ in Eq.~(\ref{eq:flowvel_TG2}) are $\sqrt{2}$ times larger than those in Eq.~(\ref{eq:flowvel_TG1}).
The length of the unit cell is $2\pi L$ in the $x_1$ and $x_2$-directions and
the flow is maintained by the external force:
\begin{eqnarray}
h_{1}= \frac{2A\nu}{L^2}\sin\left(\frac{x_1}{L}\right)\cos\left(\frac{x_2}{L}\right),\;\;\;
h_{2}=-\frac{2A\nu}{L^2}\cos\left(\frac{x_1}{L}\right)\sin\left(\frac{x_2}{L}\right).
\label{eq:flowext_TG2}
\end{eqnarray}
As in the previous section, we compare particle trajectories and  
the induced disturbance velocity field obtained by different numerical models, 
with particular emphasis on investigating the history effect on the trajectory for different initial particle positions.
Finally, the flow-mediated interaction between multiple particles at distances around $10D$ is studied.
All the results in this section are obtained for density ratio $\rho_d/\rho_c=100$
and the Reynolds number ${\rm Re}=AL/\nu=30$ (corresponding to ${\rm Re}=15$ 
with the definition in Sec.~\ref{sec:result_vor_nograv1}).
The time step is $(A/L)\Delta t=5.52\times 10^{-4}$ for the fully-resolved simulation
and $(A/L)\Delta t=1.10\times 10^{-2}$ in the other cases.
Gravity works in the negative $x_2$-direction.
For all the simulations, the initial particle velocity is the same as the flow at the particle position
and the angular velocity is 0.

\subsubsection{Validation of the VA simulation}\label{sec:result_vor_grav1_comparisonibm}
The particle initial position is $(x_1/L,x_2/L,x_3/L)=(\pi/2,\pi/2,0)$
where the flow velocity is 0.
The particle diameter is $D/L=2\pi/16$ (${\rm St}=12.9$) and the Galileo number is ${\rm Ga}=8.44$.
The domain size is $l_1/L=4\pi$ and $l_2/L=l_3/L=2\pi$.
The number of grid points is $768\times 384\times 384$ for the fully-resolved IBM simulation
and $64\times 32\times 32$ for the VA and the TT simulations.

\begin{figure}[!h]
\centering
 \psfrag{x}[s][s][1]{$x_1/\pi L$}
 \psfrag{y}[c][c][1]{$x_2/\pi L$}
\begin{overpic}[scale=0.7, angle=-90]{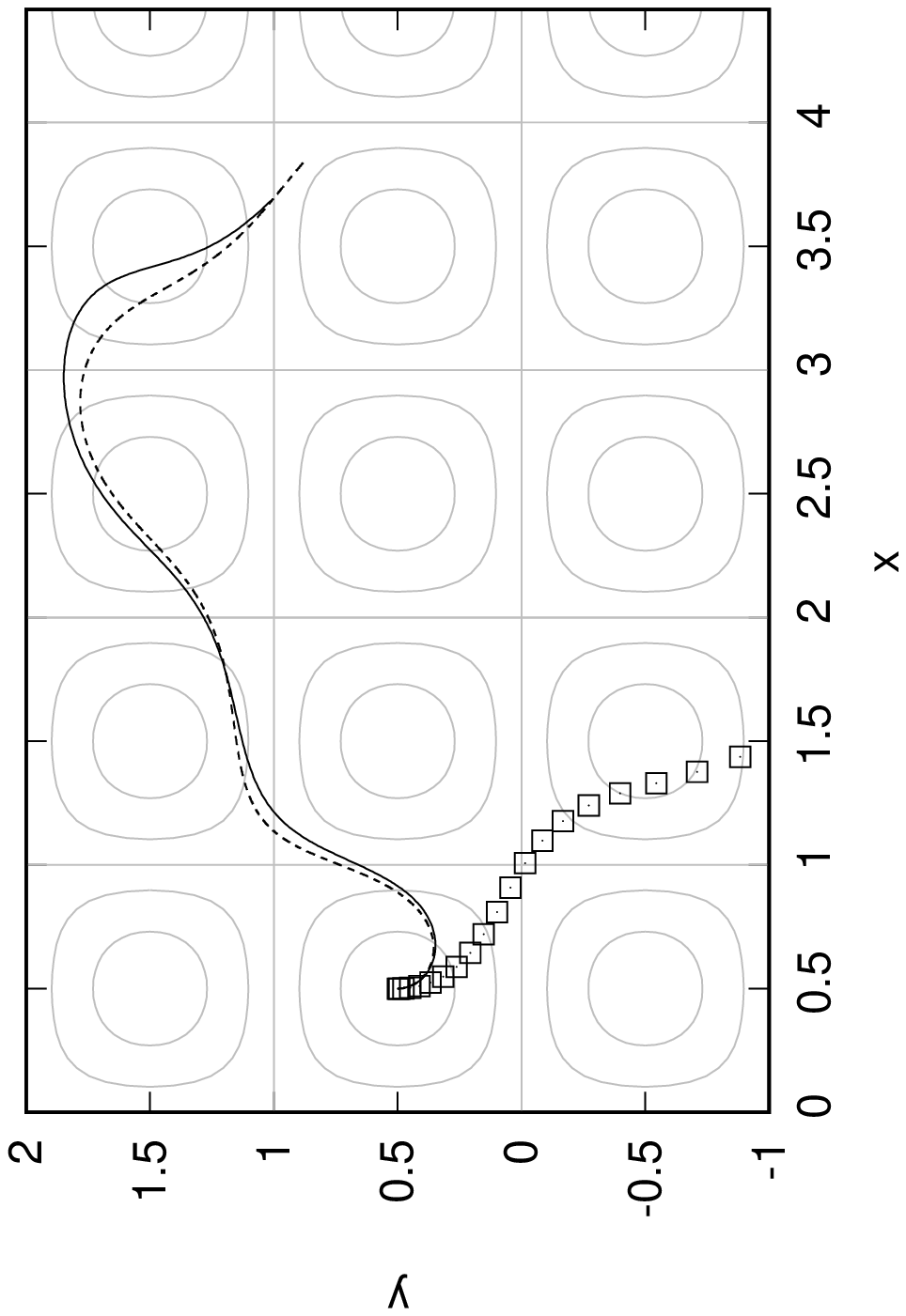}
  \put(0,65){$(a)$}
\end{overpic}\\
\begin{overpic}[scale=0.7, angle=-90]{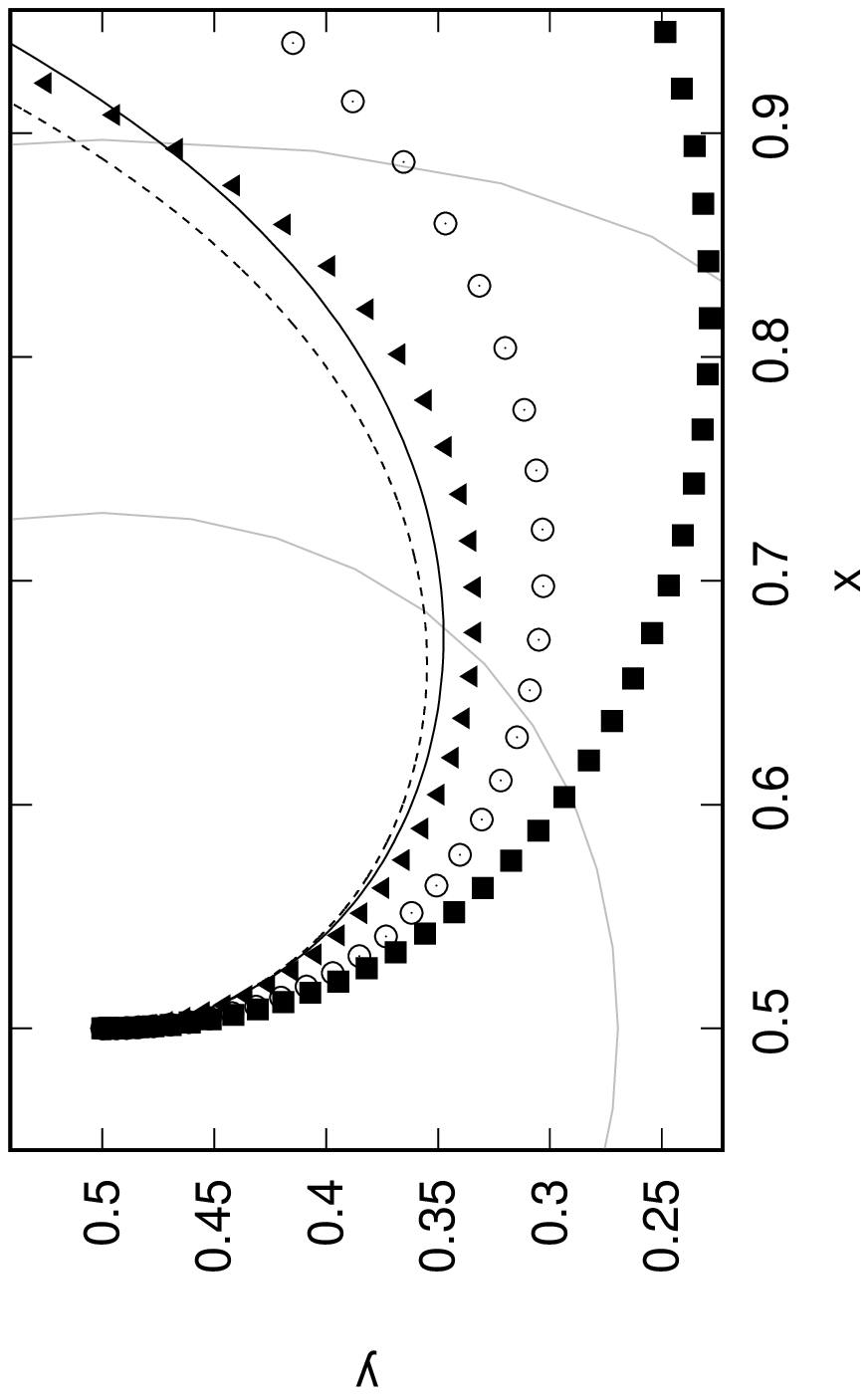}
  \put(0,65){$(b)$}
\end{overpic}
\caption{(a) Trajectory of a single particle with density ratio $\rho_d/\rho_c=100$ under the gravity in an array of Taylor-Green vortices.
Solid line, VA simulation; 
dashed line, fully-resolved simulation;
open square, TT simulation;
The grey lines show the streamlines of the undisturbed flow.
(b) Enlarged view around the initial particle position.
Filled triangle,  O-LB simulation; filled square, O-L simulation; open circle, O-NL simulation.}
\label{fig:celljob_test_xy}
\end{figure}

Figure~\ref{fig:celljob_test_xy} (a) compares the particle trajectories for the different simulations.
The result of the VA simulation shows good agreement with that of the fully-resolved IBM simulation.
The trajectory is not as simple as in the no-gravity cases:
the particle is accelerated by gravity initially and then
is transported upward by the vortex.
Interestingly, the TT simulation 
does not yield the upward particle motion due to the reduced value of the drag force.
Figure~\ref{fig:celljob_test_xy} (b) compares the initial stage of the trajectories 
obtained by the different formulations including the one-way coupling regime.
The result of the O-LB simulation, Eq.~(\ref{eq:vpcl_l01}), is the closest (among the one-way coupling simulations)
to that of the fully-resolved IBM simulation.
Therefore, in the initial stage, the history effect is more important than the nonlinearity of the drag model.
However, the difference in the trajectory of the O-LB simulation increases 
with respect to the reference case after the initial stage of Fig.~\ref{fig:celljob_test_xy} (b) (i.e., when the particle goes into the neighbouring vortex) 
because of the linear drag model and the error in the Basset term at longer times.
The result of the VA simulation suggests that an appropriate
two-way coupling model can reproduce the history effect without a complicated model
when the effect of unsteady disturbance due to the finite-size particle is reflected.

The particle Reynolds number exhibits temporal variation up to around 10
with accelerated and decelerated motion (figure omitted).


\subsubsection{History effect on particle trajectories}\label{sec:result_vor_grav1_history}
To highlight the importance to include the history effect,
we investigate the particle trajectories for different initial particle positions and
two particle diameters, $D/L=2\pi/16$ (${\rm Ga}=8.44$, ${\rm St}=12.8$) and $D/L=2\pi/32$ (${\rm Ga}=2.98$, ${\rm St}=3.21$).
The domain size is $l_1/L=l_2/L=8\pi$ and $l_3/L=4\pi$ for the larger particle and
$l_1/L=l_2/L=4\pi$ and $l_3/L=2\pi$ for the smaller particle.
The number of grid points is $128\times 128\times 64$ for both cases.

\begin{figure}[!h]
\centering
 \psfrag{x}[s][s][1]{$x_1/\pi L$}
 \psfrag{y}[c][c][1]{$x_2/\pi L$}
\begin{overpic}[scale=0.48, angle=-90]{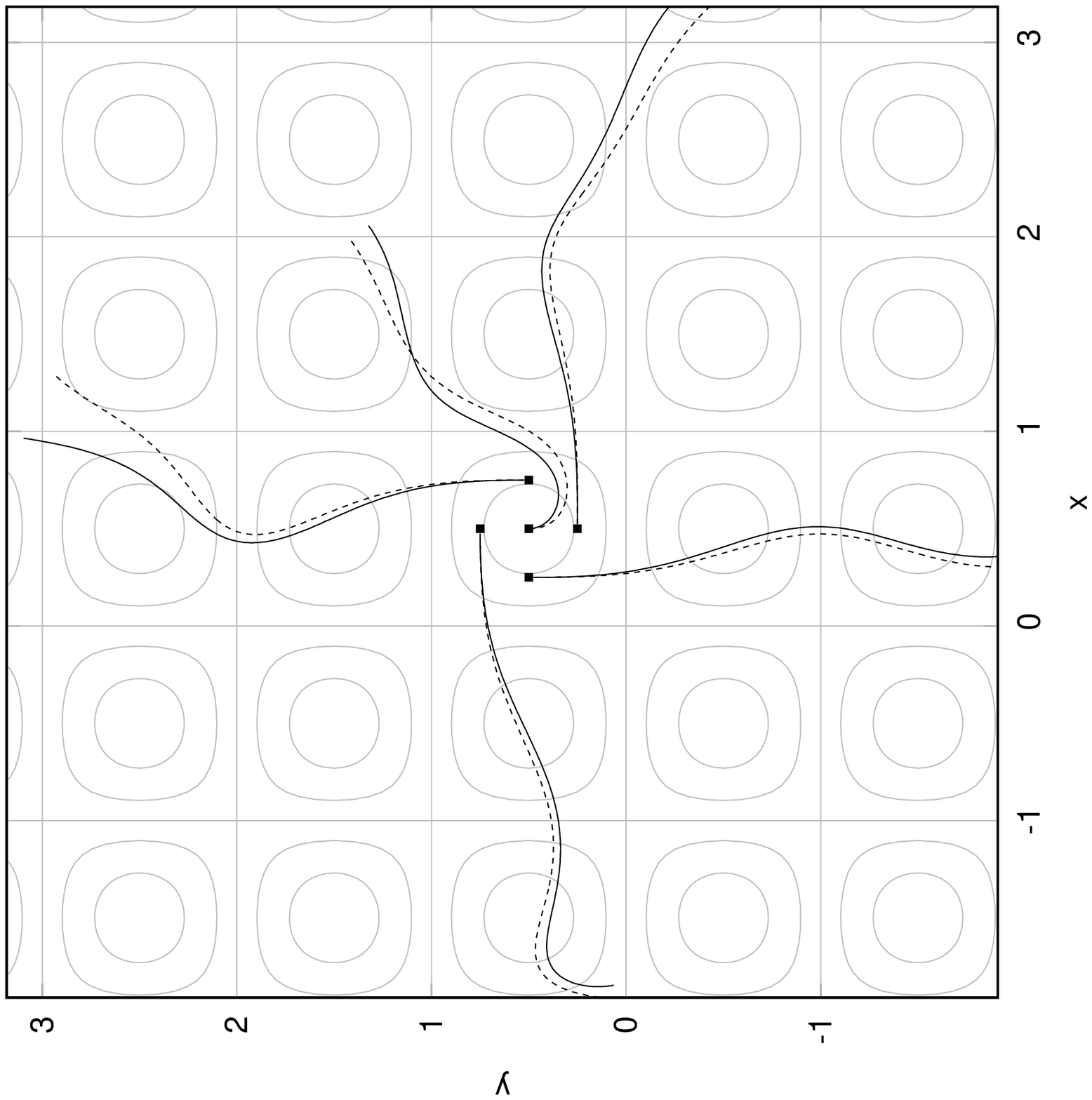}
  \put(10,65){$(a)$}
\end{overpic}\\
\begin{overpic}[scale=0.48, angle=-90]{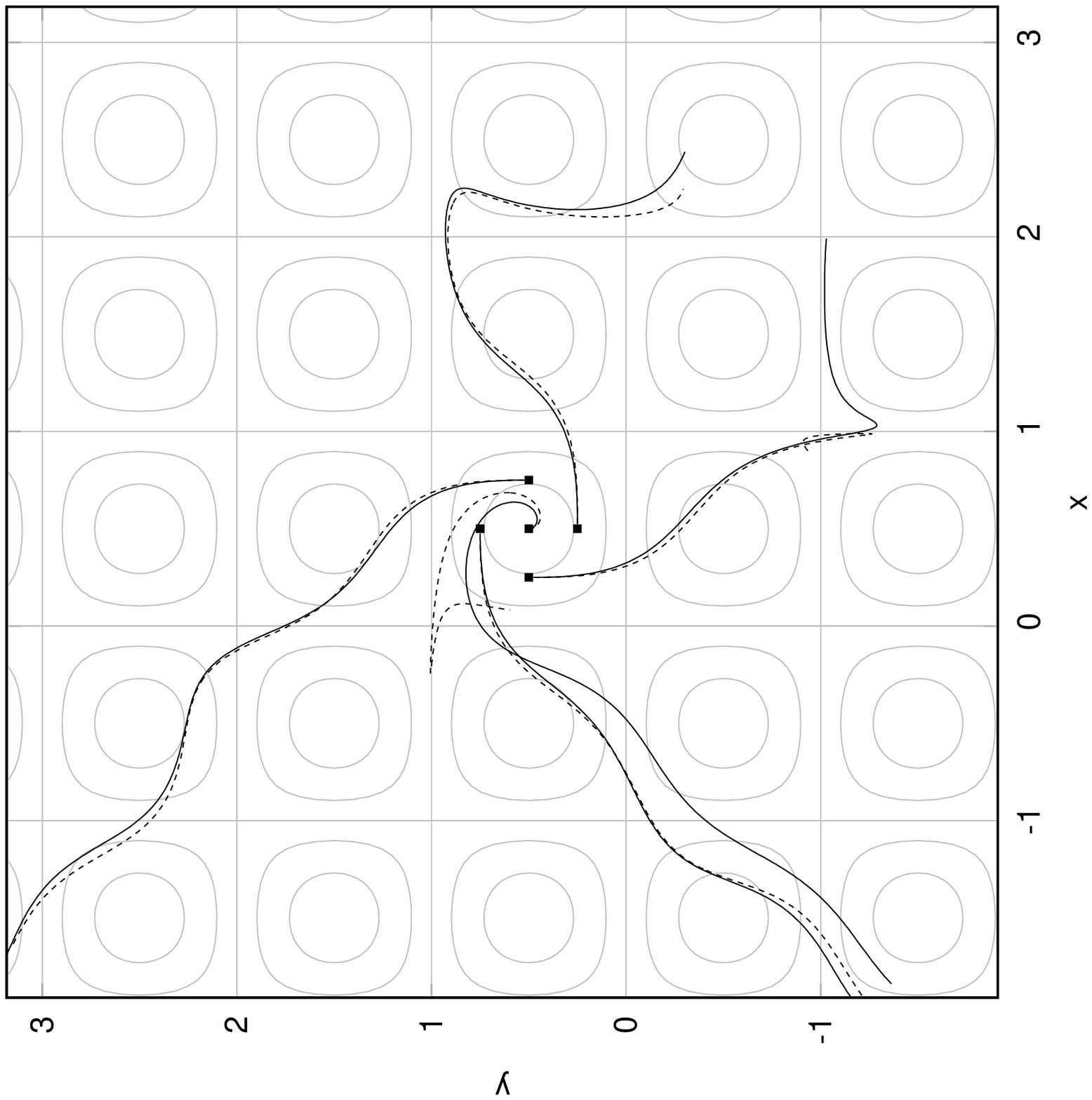}
  \put(10,65){$(b)$}
\end{overpic}
\caption{Trajectories of particles in an array of Taylor-Green vortices under the gravity of different size (a) $D/L=2\pi/16$ and (b) $D/L=2\pi/32$ and different initial positions, indicated by the filled squares. Solid line, VA simulation; dashed line, O-NL simulation.
The grey lines show the streamlines of the undisturbed flow.}
\label{fig:celljob_dd_xy}
\end{figure}

The trajectories pertaining five different initial particle positions
(along an enclosed streamline and at the vortex centre),
$(x_1/L,x_2/L)=(\pi/2,\pi/2)$, $(\pi/4,\pi/2)$, $(3\pi/4,\pi/2)$, $(\pi/2,\pi/4)$ and $(\pi/2,3\pi/4)$, 
are displayed in Fig.~\ref{fig:celljob_dd_xy}.
The trajectories are obtained with the VA approach and the O-NL simulation excluding the history effect.
The trajectories obtained with the VA simulation have slightly larger curvature than those from the O-NL simulation
at the early stage, which is consistent with the observations above about the role of the history effects.
The distances between the corresponding trajectories increase with time. 
For the larger particle (Fig.~\ref{fig:celljob_dd_xy} (a)),
the differences are already non-negligible in the cell adjacent to that of the initial particle positions.
For the smaller particle (Fig.~\ref{fig:celljob_dd_xy} (b)),
except for the particle with the initial position $(x_1/L,x_2/L)=(\pi/2,\pi/2)$,
the differences between the two models are relatively small.
This trend is explained by the fact that 
the history effect becomes smaller for smaller particles (Bergougnoux et al., 2014; Daitche, 2015).
For the case with the initial position $(x_1/L,x_2/L)=(\pi/2,\pi/2)$,
the long-time less-active motion around the vortex centre enhances the history effect on the trajectory.

\subsubsection{Interaction between multiple particles}\label{sec:result_vor_grav1_threept}
\begin{figure}[!h]
\centering
 \psfrag{x}[s][s][1]{$x_1/\pi L$}
 \psfrag{y}[c][c][1]{$x_2/\pi L$}
\begin{overpic}[scale=0.5, angle=-90]{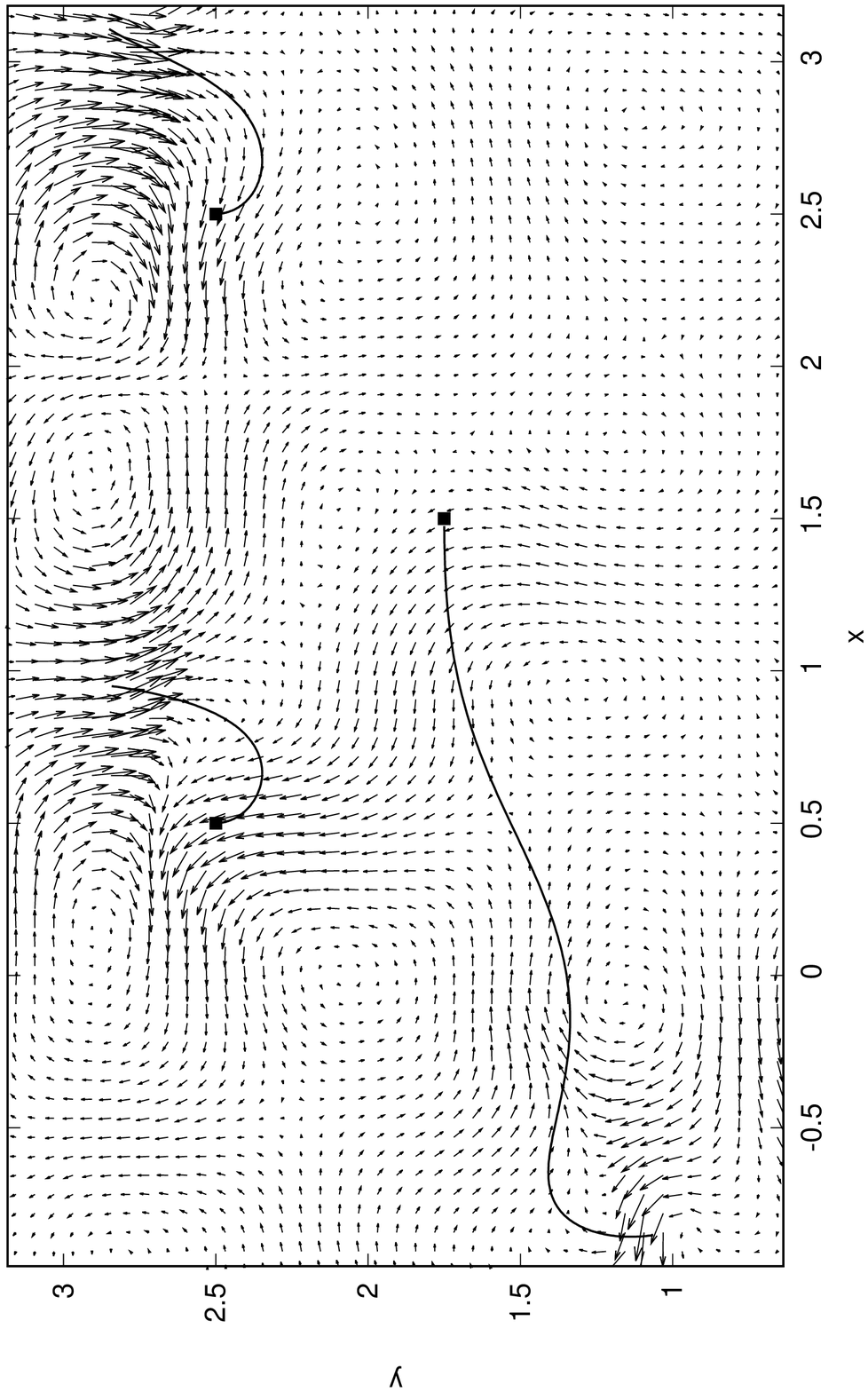}
  \put(-2,60){$(a)$}
\end{overpic}
\begin{overpic}[scale=0.5, angle=-90]{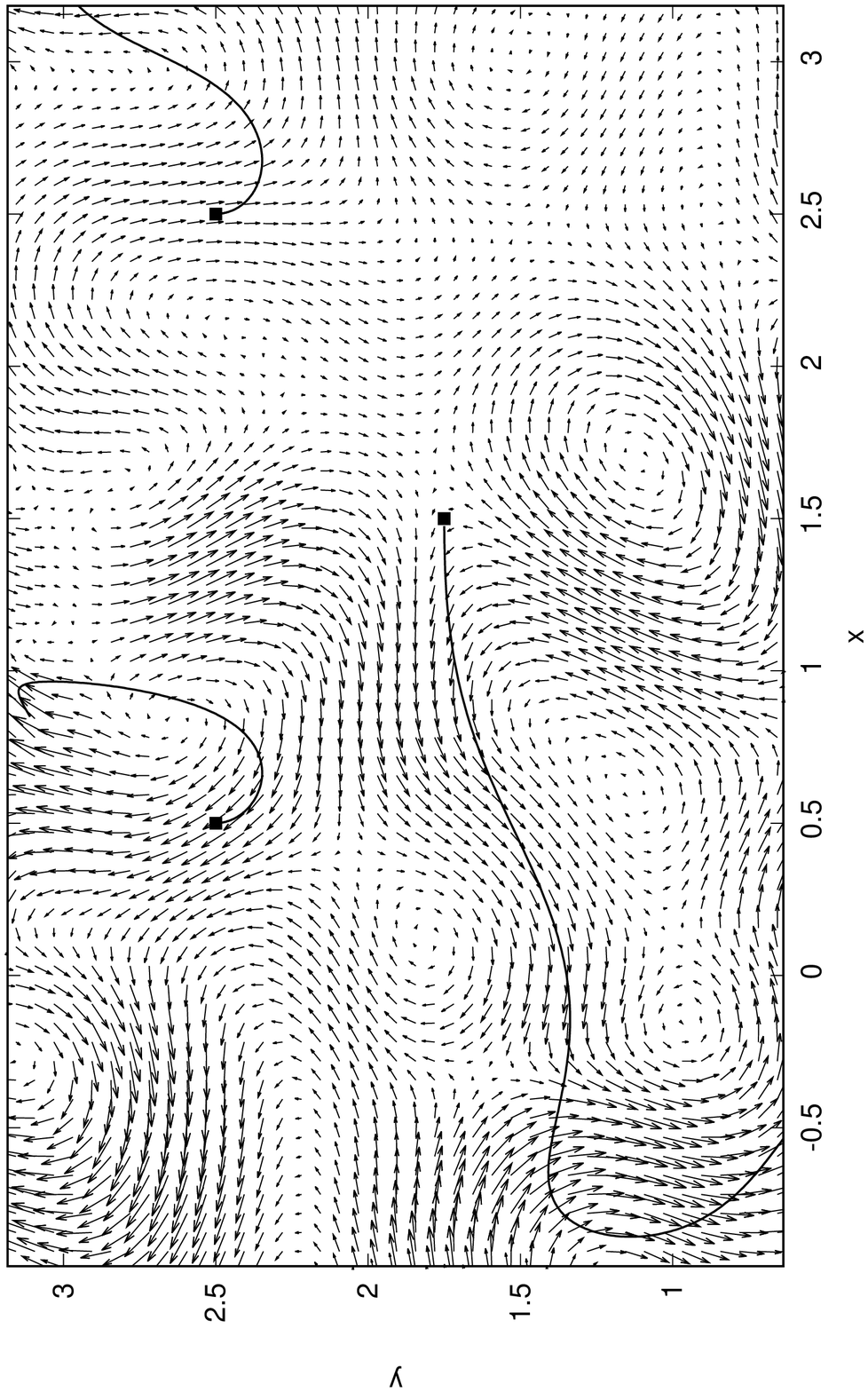}
  \put(-2,60){$(b)$}
\end{overpic}
\caption{Disturbance flow field and particle trajectories at 
(a) $(A/L)t=22.08$ and (b) $(A/L)t=33.12$ for three particles released in an array of Taylor-Green vortices under the gravity.
The square symbols indicate the initial particle positions.}
\label{fig:celljob_pt3case4_vec}
\end{figure}

The importance of the two-way coupling simulation for the inter-particle interaction
through the flow disturbance is demonstrated in the following.
For the simulations presented here, the physical parameters are $D/L=2\pi/16$ and ${\rm Ga}=8.44$.
The domain size is $l_1/L=l_2/L=8\pi$ and $l_3/L=4\pi$ and
the number of grid points $128\times 128\times 64$.
The interaction between particles at distances of around $10D$, 
which is a typical distance for volume fraction $\mathcal{O}(10^{-4})$,
is simulated with 3 particles with initial positions 
$(x_1/L,x_2/L,x_3/L)=(\pi/2,5\pi/2,0)$, $(5\pi/2,5\pi/2,0)$ and $(3\pi/2,3\pi/2,0)$.
Note that the first two particles are in the same relative position of the 
respective Taylor-Green vortex units.
Figure~\ref{fig:celljob_pt3case4_vec} shows the disturbance flow field and the particle trajectories
at two different instants.
If the inter-particle interactions are ignored, the trajectories of the two particles
initially at $x_2/L=5\pi/2$ should be the same.
As shown in Fig.~\ref{fig:celljob_pt3case4_vec} (b) at time $(A/L)t=33.12$, however,
the trajectory of the particle released from $(x_1/L,x_2/L,x_3/L)=(\pi/2,5\pi/2,0)$ 
turns to a different direction in comparison to that released from $(x_1/L,x_2/L,x_3/L)=(5\pi/2,5\pi/2,0)$.
The present result suggests that the particle motion for $D\sim\eta$ is clearly influenced 
by other particles at distance around $10D$.
Also, the flow disturbance around $(x_1/L,x_2/L,x_3/L)=(\pi/2,5\pi/2,0)$ is larger than 
that around $(x_1/L,x_2/L,x_3/L)=(5\pi/2,5\pi/2,0)$ owing to the inter-particle interaction.
The spreading of the disturbance velocity over a wide region is caused by convection 
since the convective time scale $(L/A)$ is sufficiently smaller than 
the viscous time scale $(L^2/\nu)$ (i.e., $\nu/AL=1/30$) in our case. 
As the modeling of the convective effect is difficult,
the two-way coupling simulation is necessary to investigate the inter-particle interaction.

\section{Conclusion}
For the simulation of flows laden with particles of size comparable to the smallest turbulent eddies, $D\sim\eta$, 
we have previously developed an  interaction force model based on the volume-averaged 
continuity and momentum equations.
In this paper, we proposed a new method to  estimate 
the fluid force to enable simulation of the transport of particles within the same volume-averaged framework (VA simulation). 
The VA velocity at the particle centre is correlated with the particle Reynolds number.
At the same time, the effects of the pressure gradient, the velocity gradients and the particle rotation 
are incorporated into the interaction force model.
The qualitative advantages of the VA approach are 
the capability of representing the history effect without a complicated model
and the better drag estimation compared to the traditional point-model based method.

To test the proposed model, we set up configurations of increasing complexity
and compared the results with those obtained with interface-resolved simulations based on the immersed-boundary method (IBM).
When considering a single settling particle in a stationary fluid,
we showed that the history effect was captured in the VA simulation without any specific model.
We then examined the flow around a rotating particle at ${\rm Re}_p\leq 20$ and showed that
the direction of the lift force was represented by the model, 
and the magnitudes for ${\rm Re}_p< 10$ agreed
with those in other studies (Bagchi and Balachandar, 2002; Bluemink et al., 2010).
Therefore, in the present cases, the proposed drag estimation method reflects the disturbance flow
that contributes to the history and the lift forces.

To show the applicability of VA simulation for the study of turbulent modulation, 
the simulation for the Taylor-Green vortex at Reynolds numbers ${\rm Re}=15$ and $18$ was carried out
with the particle diameter being $\mathcal{O}(10)$ times smaller than the vortex.
For density ratio $\rho_d/\rho_c\geq 10$, 
the particle motion obtained by the VA simulation showed much better agreement
with that of the fully-resolved simulation than the traditional two-way coupling simulation.
The disturbance flow 
also showed good agreement with that of the fully-resolved simulation.
On the other hand, for density ratio $\rho_d/\rho_c=1$,
the VA simulation model needs to be improved.
For a further improvement of the estimation of the fluid force,
unsteadiness and non-uniformity of the flow need to be considered.
However, we consider the method as promising as the computational cost of the VA simulation is $\mathcal{O}(10^4)$ times lower than that of the 
fully-resolved IBM simulation in the present paper.

The importance of two-way coupling for the correct prediction of particle trajectories in vortical flows 
was confirmed for $\rho_d/\rho_c=100$.
For particles released in a vortical array, 
the trajectory curvature in the initial stage increased due to the history effects, which clearly influenced the future dynamics.
The history effect estimated in the VA simulation
tends to be larger for larger particle as supported by Bergougnoux et al. (2014) and Daitche (2015).
For particles initially placed at the vortex centre,
the long residence time around the initial position increases the importance of the role of the history effects on the trajectory.
It is also found that the particle interactions, assuming an average inter-particle distance of about $10D$,
influence the particle motion in vortical flows.
These results suggest that the history effects and inter-particle flow-mediated interactions need to be considered
by two-way coupling simulations even in dilute particle-laden turbulence.

\section*{Acknowledgements}
One of the authors, T.F., gratefully acknowledges the financial support of 
the Japan Society for the Promotion of Science (JSPS) KAKENHI Grant No.~15J00439. 
L.B. and W.F.  acknowledge financial support from the European Research Council grant no.\ ERC-2013-
CoG-616186, TRITOS, by the Swedish Research Council (VR) and computer time
provided by SNIC (Swedish National Infrastructure for Computing).
This work is partly supported by Grant-in-Aid (B) No.16H04271 and No.17H03174 of the JSPS.
The authors gratefully acknowledge the financial support by Mr.~Yibin Lin (Hangzhou, China).

\renewcommand{\theequation}{A.\arabic{equation}}
\setcounter{equation}{0} 
\section*{Appendix A:~Differentiability of volume-averaged quantities}
The form of the viscous term in Eq.~(\ref{eq:VANS01}) is justified in this section.
In the following discussion, $V$ is assumed to be larger than the particle as used in this paper.
The volume integral $Q(\bm{x})$ is defined as
\begin{eqnarray}
Q(\bm{x})=\int_{V}qdV,
\end{eqnarray}
where $q$ is a bounded function defined in both fluid and solid.
We also assume that Taylor expansion of $q$ is possible except on the interface.
The volume-averaged quantities correspond to $Q(\bm{x})/V$.
For example, $\alpha_c\left<u_1\right>_c$ is constructed from
\begin{eqnarray}
  q=
   \begin{cases}
     \;  u_1
     \;\;\;{\rm inside~the~fluid}\\
     \; 0
     \;\;\;\;\;{\rm inside~the~solid}
   \end{cases},
  \label{eq:app_ex1ave}
\end{eqnarray}
and $(\alpha_c\left<u_1\right>_c+\alpha_d\left<v_1\right>_d)$ corresponds to 
\begin{eqnarray}
  q=
   \begin{cases}
     \;  u_1
     \;\;\;{\rm inside~the~fluid}\\
     \; v_1
     \;\;\;\:{\rm inside~the~solid}
   \end{cases}.
  \label{eq:app_ex2ave}
\end{eqnarray}
In the following, the first and second-order derivatives of $Q(\bm{x})$ are considered.

Figure~\ref{fig:app_diff1} 
schematically shows the geometric relation between 
$V(\bm{x})$ and $V(\bm{x}+h\bm{e}_i)$, 
and we focus on the integration of $q$ over the volume denoted as $Q(\bm{x})$ and $Q(\bm{x}+h\bm{e}_i)$.
The volume integrals of $q$ in the shaded regions are denoted as $Q^{+}$, $Q^{-}$ and $Q_{\rm cut}$.
The surface of $V(\bm{x})$ is denoted as $S$.
The outward unit normal vector on $S$ is denoted as $\bm{n}_V$.
The surface $S$ is divided into a region denoted as $S^{+}$, where $\bm{n}_V\cdot\bm{e}_i\geq 0$, 
and $S^{-}$, defined by $\bm{n}_V\cdot\bm{e}_i\leq 0$.
According to Fig.~\ref{fig:app_diff1},
the volume integrals are
\begin{eqnarray}
Q(\bm{x}+h\bm{e}_i)&=&Q(\bm{x})+Q^+-Q^-,\\
Q^+&=&\int_{S^+} \bm{n}_V\cdot\bm{e}_i\int_0^h q(\bm{S}+l\bm{e}_i)dldS-Q_{\rm cut},\\
Q^-&=&\int_{S^-}-\bm{n}_V\cdot\bm{e}_i\int_0^h q(\bm{S}+l\bm{e}_i)dldS-Q_{\rm cut},
\end{eqnarray}
where $\bm{S}$ indicates the position on $S$.
Therefore, we obtain the following equation:
\begin{eqnarray}
\frac{Q(\bm{x}+h\bm{e}_i)-Q(\bm{x})}{h}&=&\frac{Q^+-Q^-}{h}\nonumber\\
&=&\frac{1}{h}\int_{S}\bm{n}_V\cdot\bm{e}_i\int_0^h q(\bm{S}+l\bm{e}_i)dl\, dS.
\end{eqnarray}
To deal with the interface between the phases, 
we define the fraction of the surface $S_{\rm jump} (\subset S)$ such that
\begin{eqnarray}
S_{\rm jump}=\left\{\bm{S}{\rm\;}|{\rm\;}\bm{S}+\beta h\bm{e}_i\in
{\rm particle\; interface\;\;}(0\leq \beta\leq 1)\right\},
\nonumber
\end{eqnarray}
(see Fig.~\ref{fig:app_diff1}).
With this decomposition of the surface $S$,
\begin{eqnarray}
\frac{Q(\bm{x}+h\bm{e}_i)-Q(\bm{x})}{h}
&=&\int_{S-S_{\rm jump}}\big\{q(\bm{S})+O(h)\big\}\bm{n}_V\cdot\bm{e}_idS\nonumber\\
&&+\int_{S_{\rm jump}}\bm{n}_V\cdot\bm{e}_i\frac{1}{h}\int_0^h q(\bm{S}+l\bm{e}_i)dl \, dS.
\end{eqnarray}
Taking the limit of $h\to 0$, $S_{\rm jump}$ converges to 0 and 
\begin{eqnarray}
\frac{1}{h}\int_0^h q(\bm{S}+l\bm{e}_i)dl \nonumber
\end{eqnarray}
on $S_{\rm jump}$ is bounded since $q$ is bounded.
We can therefore write
\begin{eqnarray}
\lim_{h\to 0}\frac{Q(\bm{x}+h\bm{e}_i)-Q(\bm{x})}{h}
&=&\int_{S}q(\bm{S})\bm{n}_V\cdot\bm{e}_idS.
\label{eq:app_avediff01a}
\end{eqnarray}
As Eq.~(\ref{eq:app_avediff01a}) holds regardless of the sign of $h$,
we obtain the derivative
\begin{eqnarray}
\frac{\partial Q(\bm{x})}{\partial x_i}
=\int_{S}q(\bm{S})\bm{n}_V\cdot\bm{e}_idS.
\label{eq:app_avediff01b}
\end{eqnarray}
Note that $S_{\rm jump}\to 0$ is guaranteed by the size difference between $V$ and the particle.
In this case, the continuity of $q$ on the interface is not necessary for the first-order derivative.

For simplicity, ${\partial Q(\bm{x})}/{\partial x_i}$ is denoted as $Q_{,i}(\bm{x})$.
For higher-order derivatives, we consider that
\begin{eqnarray}
Q_{,j}(\bm{x}+h\bm{e}_i)
&=&\int_{S}q(\bm{S}+h\bm{e}_i)\bm{n}_V\cdot\bm{e}_jdS\nonumber\\
&=&\int_{S-S_{\rm jump}}\left\{q(\bm{S})+\frac{\partial q}{\partial x_i}(\bm{S})h+O(h^2)\right\}
\bm{n}_V\cdot\bm{e}_jdS\nonumber\\
&&+\int_{S_{\rm jump}}q(\bm{S}+h\bm{e}_i)\bm{n}_V\cdot\bm{e}_jdS,
\end{eqnarray}
so that
\begin{eqnarray}
\frac{Q_{,j}(\bm{x}+h\bm{e}_i)-Q_{,j}(\bm{x})}{h}
&=&\int_{S-S_{\rm jump}}\left\{\frac{\partial q}{\partial x_i}(\bm{S})+O(h)\right\}\bm{n}_V\cdot\bm{e}_jdS\nonumber\\
&&+\int_{S_{\rm jump}}\left\{\frac{q(\bm{S}+h\bm{e}_i)-q(\bm{S})}{h}\right\}\bm{n}_V\cdot\bm{e}_jdS.
\label{eq:app_avediff02a}
\end{eqnarray}
By denoting the interface as $(\bm{S}+k(\bm{S})\bm{e}_i)$ with $0\leq k\leq h$,
we obtain
\begin{eqnarray}
\frac{q(\bm{S}+h\bm{e}_i)-q(\bm{S})}{h}=\frac{[q(\bm{S}+k\bm{e}_i)]}{h}+O(h^0),
\end{eqnarray}
where $[\cdot]$ represents the jump of the function at the interface.
Taking the limit of $h\to 0$, Eq.~(\ref{eq:app_avediff02a}) yields:
\begin{eqnarray}
\lim_{h\to 0}\frac{Q_{,j}(\bm{x}+h\bm{e}_i)-Q_{,j}(\bm{x})}{h}
&=&\int_{S}\frac{\partial q}{\partial x_i}(\bm{S})\bm{n}_V\cdot\bm{e}_jdS\nonumber\\
&&+\lim_{h\to 0}\int_{S_{\rm jump}}\frac{[q(\bm{S}+k\bm{e}_i)]}{h}\bm{n}_V\cdot\bm{e}_jdS.
\label{eq:app_avediff02b}
\end{eqnarray}
In general, the right-hand side of Eq.~(\ref{eq:app_avediff02b}) depends on the sign of $h$
(e.g., the second-order derivative of $\alpha_d$ is not determined).
On the other hand, when $[q(\bm{S}+k\bm{e}_i)]=0$, 
we can define the second-order derivative as
\begin{eqnarray}
\frac{\partial^2 Q(\bm{x})}{\partial x_i\partial x_j}=\int_{S}\frac{\partial q}{\partial x_i}(\bm{S})\bm{n}_V\cdot\bm{e}_jdS.
\label{eq:app_avediff02c}
\end{eqnarray}
As the velocities $\bm{u}$ and $\bm{v}$ are continuous across the interface,
the viscous term $\nabla^2(\alpha_c\left<\bm{u}\right>_c+\alpha_d\left<\bm{v}\right>_d)$ is well-defined. Note, however, that 
the decomposition into $\nabla^2(\alpha_c\left<\bm{u}\right>_c)+\nabla^2(\alpha_d\left<\bm{v}\right>_d)$ is not allowed.

\renewcommand{\theequation}{B.\arabic{equation}}
\setcounter{equation}{0} 
\section*{Appendix B:~Calculation of interaction force}
The interaction force can be written as
\begin{eqnarray}
\bm{f}=-\int_{S_d}\frac{1}{\rho_c}\bm{s}\cdot\bm{n}dS,
\end{eqnarray}
where $\bm{s}$ is the stress on the surface:
\begin{eqnarray}
\bm{s}=-\delta p\bm{I}
+\rho_c\nu\left(\nabla\bm{u}+(\nabla\bm{u})^T\right).
\end{eqnarray}
In Sec.~\ref{sec:intf_models}, the stress vectors:
\begin{eqnarray}
  \bm{s}\cdot\bm{n}=\rho_c\frac{3(\bm{F}_{\rm pg}\cdot\bm{n})}{\pi D^2}\bm{n}
\end{eqnarray}
and
\begin{eqnarray}
  \bm{s}\cdot\bm{n}=-3\rho_c\nu\bm{\Omega}_p\times\bm{n}
\end{eqnarray}
are used for the modeling of Eqs.~(\ref{eq:intf_pg01}) and (\ref{eq:intf_rot01}).
Therefore, the following two integrals are enough 
for the derivation of the interaction force models:
\begin{eqnarray}
  \int_{S_d}n_idS,\;\;\;\int_{S_d}n_in_jdS.
\end{eqnarray}
Note that these integrals are also enough for the interaction force modelling
in Fukada et al. (2016).

We consider three unit vectors $\bm{A}$, $\bm{B}$ and $\bm{C}$ with
$\bm{A}={(\bm{x}-\bm{x}_p)}/{|\bm{x}-\bm{x}_p|}$ and 
$\bm{A}\cdot\bm{B}=\bm{B}\cdot\bm{C}=\bm{C}\cdot\bm{A}=0$.
The basis vector $\bm{e}_i$ and $\bm{n}$ can be written as
\begin{eqnarray}
  \bm{e}_i&=&(\bm{A}\cdot\bm{e}_i)\bm{A}+(\bm{B}\cdot\bm{e}_i)\bm{B}+(\bm{C}\cdot\bm{e}_i)\bm{C},\\
  \bm{n}&=&(\bm{A}\cdot\bm{n})\bm{A}+(\bm{B}\cdot\bm{n})\bm{B}+(\bm{C}\cdot\bm{n})\bm{C},
\end{eqnarray}
so that the integrands become
\begin{eqnarray}
  n_i&=&\bm{n}\cdot\bm{e}_i\nonumber\\
  &=&(\bm{A}\cdot\bm{e}_i)(\bm{A}\cdot\bm{n})
  +(\bm{B}\cdot\bm{e}_i)(\bm{B}\cdot\bm{n})+(\bm{C}\cdot\bm{e}_i)(\bm{C}\cdot\bm{n}),\\
  n_in_j&=&(\bm{A}\cdot\bm{e}_i)(\bm{A}\cdot\bm{e}_j)(\bm{A}\cdot\bm{n})^2
  +(\bm{B}\cdot\bm{e}_i)(\bm{B}\cdot\bm{e}_j)(\bm{B}\cdot\bm{n})^2
  +(\bm{C}\cdot\bm{e}_i)(\bm{C}\cdot\bm{e}_j)(\bm{C}\cdot\bm{n})^2
  \nonumber\\
  &&+\left\{(\bm{A}\cdot\bm{e}_i)(\bm{B}\cdot\bm{e}_j)
  +(\bm{B}\cdot\bm{e}_i)(\bm{A}\cdot\bm{e}_j)\right\}
  (\bm{A}\cdot\bm{n})(\bm{B}\cdot\bm{n})
  \nonumber\\
  &&+\left\{(\bm{B}\cdot\bm{e}_i)(\bm{C}\cdot\bm{e}_j)
  +(\bm{C}\cdot\bm{e}_i)(\bm{B}\cdot\bm{e}_j)\right\}
  (\bm{B}\cdot\bm{n})(\bm{C}\cdot\bm{n})
  \nonumber\\
  &&+\left\{(\bm{C}\cdot\bm{e}_i)(\bm{A}\cdot\bm{e}_j)
  +(\bm{A}\cdot\bm{e}_i)(\bm{C}\cdot\bm{e}_j)\right\}
  (\bm{C}\cdot\bm{n})(\bm{A}\cdot\bm{n}).
\end{eqnarray}
Given the symmetry about the direction $\bm{A}$, the integrals reduce to 
\begin{eqnarray}
  \int_{S_d}n_idS&=&(\bm{A}\cdot\bm{e}_i)\int_{S_d}\cos tdS,\\
  \int_{S_d}n_in_jdS&=&
  (\bm{A}\cdot\bm{e}_i)(\bm{A}\cdot\bm{e}_j)
  \int_{S_d}\cos^2 tdS
  \nonumber\\
  &&+\left\{\delta_{ij}-(\bm{A}\cdot\bm{e}_i)(\bm{A}\cdot\bm{e}_j)\right\}
  \frac{1}{2}\int_{S_d}(1-\cos^2t)dS,
\end{eqnarray}
where $t$ is the angle between $\bm{A}$ and $\bm{n}$.
In this derivation, we used the following relations:
\begin{eqnarray}
  \int_{S_d}(\bm{B}\cdot\bm{n})^2dS=\int_{S_d}(\bm{C}\cdot\bm{n})^2dS&=&
  \frac{1}{2}\int_{S_d}\{1-(\bm{A}\cdot\bm{n})^2\}dS,\\
  (\bm{B}\cdot\bm{e}_i)(\bm{B}\cdot\bm{e}_j)+(\bm{C}\cdot\bm{e}_i)(\bm{C}\cdot\bm{e}_j)
  &=&(\bm{I}-\bm{A}\bm{A})\colon\bm{e}_i\bm{e}_j \nonumber\\
  &=&\delta_{ij}-(\bm{A}\cdot\bm{e}_i)(\bm{A}\cdot\bm{e}_j).
\end{eqnarray}
According to Eqs.~(\ref{eq:appgeom_cos01}) and (\ref{eq:appgeom_cos02}) 
(shown later), we obtain
\begin{eqnarray}
  \int_{S_d}n_idS&=&(\bm{A}\cdot\bm{e}_i)\pi D^2\xi(1-\xi),\\
  \int_{S_d}n_in_jdS&=&(\bm{A}\cdot\bm{e}_i)(\bm{A}\cdot\bm{e}_j)
  \pi D^2\xi(1-\xi)(1-2\xi)
  +\delta_{ij}\frac{\pi D^2}{3}\xi^2(3-2\xi).
\end{eqnarray}

\renewcommand{\theequation}{C.\arabic{equation}}
\setcounter{equation}{0} 
\section*{Appendix C:~Estimation of fluid force}
The applicability of the current approximation of the fluid force is tested for unsteady flows.
For the flow fields around the particle obtained by the fully-resolved simulation,
the contributions of $F_{\rm drag}$ and $\nabla P_{\rm ud}$ on the particle acceleration (\ref{eq:vpcl01})
are computed by Eqs.~(\ref{eq:fdragnl}), (\ref{eq:rep_estimate01}) and (\ref{eq:pg_estimate01}).
Instead, the volume-averaged values are directly computed from the flow fields.
In particular, we consider time $(A/L)t=33.13$ for the cases of $\rho_d/\rho_c=1,10,1000$ in Sec.~\ref{sec:result_vor_nograv1}
and time $(A/L)t=55.22$ for the case  $\rho_d/\rho_c=100$ in Sec.~\ref{sec:result_vor_grav1_comparisonibm}.
Table~\ref{tab:result_force_estimate} shows the following three dimensionless accelerations (for comparing the contribution of each term to the right-hand side of
Eq.~(\ref{eq:vpcl01})):
\begin{eqnarray}
\bm{a}^{sv}&=&\frac{2}{2m_d+m_c}\frac{D^3}{\nu^2}F_{\rm drag}\bm{m},
\\
\bm{a}^{sv+pg}&=&\frac{2}{2m_d+m_c}\frac{D^3}{\nu^2}\left\{F_{\rm drag}\bm{m}
-\frac{\pi D^3}{4}\nabla P_{\rm ud}\right\},
\\
\bm{a}^{FR}&=&\frac{D^3}{\nu^2}\left\{\frac{d\bm{v}_p}{dt}
-\frac{2}{2m_d+m_c}\left[(m_d-m_c)\bm{g}+m_c\bm{h}\right]\right\},
\end{eqnarray}
where $d\bm{v}_p/dt$ is the net acceleration obtained from the fully-resolved IBM simulation.
Note that $\bm{a}^{sv+pg}$ is equal to $\bm{a}^{FR}$ when the errors in both the model~(\ref{eq:vpcl01}) 
and the estimation of each term are ignored.
Especially for $\rho_d/\rho_c\leq 100$, the effect of the pressure gradient is reasonably captured
by Eq.~(\ref{eq:pg_estimate01}) as $\bm{a}^{sv+pg}$ is considerably improved from $\bm{a}^{sv}$.
However, as indicated by the differences between $\bm{a}^{sv+pg}$ and $\bm{a}^{FR}$, 
unsteadiness and non-uniformity of the flow need to be considered to improve the total estimation method.

\renewcommand{\theequation}{D.\arabic{equation}}
\setcounter{equation}{0} 
\section*{Appendix D:~Calculations of geometrical functions}
The radius $R$ of the averaging volume $V$ is larger than the particle radius $r_d$.
The geometrical functions $\xi$, $\alpha_d$ and $\alpha_d\left<\bm{v}\right>_d$
are obvious for $|\bm{x}-\bm{x}_p|<R-r_d$ and $|\bm{x}-\bm{x}_p|>R+r_d$.
Therefore, we only consider $R-r_d\leq |\bm{x}-\bm{x}_p|\leq R+r_d$.
Figure~\ref{fig:app_geom1} shows the definitions of the variables considered in the following.
The origin is at the particle centre and $y$ is equal to $|\bm{x}-\bm{x}_p|$.
The variable $a$ satisfies 
\begin{eqnarray}
  r_d^2-a^2&=&R^2-(y-a)^2,\\
  a&=&\frac{r_d^2-R^2+y^2}{2y}.
  \label{eq:appgeom_a01}
\end{eqnarray}
Note that $a$ varies in the range $-r_d\leq a\leq r_d$.
The surface fraction $\xi=S_d/\pi D^2$ is 
\begin{eqnarray}
  \xi=\frac{1}{\pi D^2}\int^{\cos^{-1}(a/r_d)}_{0}(2\pi r_d\sin t) r_ddt
=\frac{1}{2}\left(1-\frac{a}{r_d}\right).
\label{eq:appgeom_xi01}
\end{eqnarray}
The volume fraction $\alpha_d=V_d/V$ is 
\begin{eqnarray}
  \alpha_d&=&\frac{1}{V}
  \left[\int^{a}_{y-R}\pi\{R^2-(y-x)^2\}dx+\int^{r_d}_{a}\pi(r_d^2-x^2)dx\right]
  \nonumber\\
  &=&\frac{y^3-3ay^2+3(a^2-R^2)y+2(r_d^3+R^3)+3a(R^2-r_d^2)}{4R^3}.
  \label{eq:appgeom_alp01}
\end{eqnarray}
The centre of gravity  $x_G$ of $V_d$ is 
\begin{eqnarray}
  x_G&=&\frac{1}{V_d}\left[\int^{a}_{y-R}\pi\{R^2-(y-x)^2\}xdx
  +\int^{r_d}_{a}\pi(r_d^2-x^2)xdx\right]\nonumber\\
  &=&\frac{1}{\alpha_d}
  \frac{y^4-6(R^2+a^2)y^2+8(R^3+a^3)y-3(R^4-r^4)+6a^2(R^2-r^2)}{16R^3}.
  \label{eq:appgeom_xg01}
\end{eqnarray}
In general, the rigid-body velocity at $\bm{x}$ can be written as 
\begin{eqnarray}
  \bm{v}&=&\bm{v}_0+\bm{\Omega}_0\times\bm{x}\nonumber\\
  &=&\bm{v}_0+\bm{\Omega}_0\times\bm{x}'+\bm{\Omega}_0\times(\bm{x}-\bm{x}'),
\end{eqnarray}
where $\bm{v}_0$ and $\bm{\Omega}_0$ are the origin velocity and 
the angular velocity around the origin.
Introducing 
\begin{eqnarray}
  \bm{x}'=\frac{1}{V_r}\int_{V_r}\bm{x}dV,
\end{eqnarray}
where $V_r$ is the volume of the rigid body,
the average velocity in the volume becomes
\begin{eqnarray}
  \frac{1}{V_r}\int_{V_r}\bm{v}dV
  =\bm{v}_0+\bm{\Omega}_0\times\bm{x}'.
\end{eqnarray}
Therefore, the averaged velocity $\alpha_d\left<\bm{v}\right>_d$ is
\begin{eqnarray}
  \alpha_d\left<\bm{v}\right>_d=\alpha_d\bm{v}_p
  +\alpha_dx_G\bm{\Omega}_p\times\frac{\bm{x}-\bm{x}_p}{|\bm{x}-\bm{x}_p|}.
  \label{eq:appgeom_xg02}
\end{eqnarray}
Eqs.~(\ref{eq:volfrac_calc}), (\ref{eq:surffrac_calc}) and (\ref{eq:volgfrac_calc}) 
are obtained by Eqs.~(\ref{eq:appgeom_xi01}),
(\ref{eq:appgeom_alp01}), (\ref{eq:appgeom_xg01}) and (\ref{eq:appgeom_xg02})
using Eq.~(\ref{eq:appgeom_a01}).

Finally, the integrals used in Appendix B are calculated as follows:
\begin{eqnarray}
  \int_{S_d}\cos tdS&=&\int_0^{\cos^{-1}(a/r_d)}2\pi r_d^2\cos t\sin t dt
  =\pi D^2\xi(1-\xi),
  \label{eq:appgeom_cos01}\\
  \int_{S_d}\cos^2 tdS&=&\int_0^{\cos^{-1}(a/r_d)}2\pi r_d^2\cos^2 t\sin t dt
  =\frac{\pi D^2}{3}\xi(4\xi^2-6\xi+3).
  \label{eq:appgeom_cos02}
\end{eqnarray}

\renewcommand{\theequation}{E.\arabic{equation}}
\setcounter{equation}{0} 
\section*{Appendix E:~Effects of size of the averaging volume and grid width}
To investigate the effect of the size of the averaging volume,
the VA simulation is expanded for the case of $R=1.5D$.
For this case, the fitting functions (\ref{eq:rep_estimate01}), (\ref{eq:fitvg_ave1}) and (\ref{eq:fitpg_ave1}) are replaced by
\begin{eqnarray}
{\rm Re}_p=1.52\left(\frac{\alpha_c|\left<\bm{u}\right>_c-\bm{v}_p|D}{\nu}\right)^{0.93},
\end{eqnarray}
\begin{eqnarray}
\delta_{urr}({\rm Re}_p)&=&-0.121\left(\frac{\nu}{D^2}\right){\rm Re}_p^{0.759},
\\[12pt]
\delta_{pr}({\rm Re}_p)&=&-0.118\left(\frac{\rho_c\nu^2}{D^3}\right){\rm Re}_p^{1.65}.
\end{eqnarray}
Based on these functions, the VA simulations under $R=1.5D$ and different $\Delta x$ are carried out for the same configurations as in
Secs.~\ref{sec:result_settling1} and \ref{sec:result_vor_grav1_comparisonibm}.

Figure~\ref{fig:app_falling_vel_R1.5} shows the result for the particle settling problem in a stationary fluid.
As for the case of $R=0.75D$, the result of the fine grid ($D/\Delta x=4$) is almost the same as that for $D/\Delta x=2$.
On the other hand, for the coarse grid ($D/\Delta x=1$), the result is quite different from the others.
Therefore, the grid resolution of $D/\Delta x=2$ is necessary to capture the averaged flow distribution around the particle.
As for the case of $R=1.5D$, the result of the coarse grid ($D/\Delta x=1$) is not so different from that of $D/\Delta x=2$.
Therefore, choosing a larger $R$ is better for $D/\Delta x=1$,
while having a smaller $R$ is appropriate for $D/\Delta x=2$.
However, to outperform the O-NL simulation (shown in Fig.~\ref{fig:falling_vel} (a)), 
a fine grid with small $R$ ($=0.75D$) is necessary.

Figure~\ref{fig:app_celljob_test_xy_R1.5} shows the result for the particle under gravity in an array of Taylor-Green vortices.
The tendency of the effects of $R$ and $\Delta x$ are quite similar to that for the settling particle in the stationary fluid.
The results for $R=1.5D$ ($D/\Delta x=1$ and 2) are still better than that of the O-NL simulation.
Therefore, for the larger $R$, the history effect is qualitatively captured in the same manner as for the smaller $R$.
The upper limit of $R$ is considered to be based on the length scale of the background flow.
In conclusion, smaller $R$ gives better results for sufficiently fine grids.
On the other hand, for coarse grid, larger $R$ is preferable.

\clearpage
\section*{References}

\begin{itemize}

\item[]
Anderson, T. B., Jackson, R. 1967
A fluid mechanical description of fluidized beds.
{\it Ind. Eng. Chem. Fundam.} {\bf 6}, pp. 527-539

\item[]
Bagchi, P., Balachandar, S. 2002
Effect of free rotation on the motion of a solid sphere in linear shear flow at moderate Re.
{\it Phys. Fluids} {\bf 14}, pp. 2719-2737

\item[]
Bagchi, P., Balachandar, S. 2003
Inertial and viscous forces on a rigid sphere in straining flows at moderate Reynolds numbers.
{\it J. Fluid. Mech.} {\bf 481}, pp. 105-148

\item[]
Bergougnoux, L., Bouchet, G., Lopez, D., Guazzelli, \'{E}. 2014
The motion of solid spherical particles falling in a cellular flow field at low Stokes number. 
{\it Phys. Fluids} {\bf 26}, 093302

\item[]
Bluemink, J. J., Lohse, D., Prosperetti, A., Wijngaarden, L. V. 2010
Drag and lift forces on particles in a rotating flow.
{\it J. Fluid. Mech.}
{\bf 643}, pp. 1-31

\item[]
Boivin, M., Simonin, O., Squires, K. D. 1998
Direct numerical simulation of turbulence modulation by particles in isotropic turbulence.
{\it J. Fluid. Mech.} {\bf 375}, pp. 235-263

\item[]
Breugem, W. P. 2012
A second-order accurate immersed boundary method for fully resolved simulations of particle-laden flows.
{\it J. Comput. Phys.} {\bf 231}, pp. 4469-4498

\item[]
Crowe, C. T., Sommerfeld, M., Tsuji, Y. 1997
{\it Multiphase Flows with Droplets and Particles,}
CRC Press

\item[]
Daitche, A. 2015
On the role of the history force for inertial particles in turbulence.
{\it J. Fluid. Mech.} {\bf 782}, pp. 567-593

\item[]
Eaton, J. K. 2009
Two-way coupled turbulence simulations of gas-particle flows using point-particle tracking.
{\it Int. J. Multiphase Flow} 
{\bf 35}, pp. 792-800

\item[]
Fornari, W., Picano, F., Brandt, L. 2016
Sedimentation of finite-size spheres in quiescent and turbulent environments.
{\it J. Fluid Mech.} {\bf 788}, pp. 640-669

\item[]
Fukada, T., Takeuchi, S., Kajishima, T. 2016
Interaction force and residual stress models for volume-averaged momentum equation 
for flow laden with particles of comparable diameter to computational grid width. 
{\it Int. J. Multiphase Flow} {\bf 85}, pp. 298-313

\item[]
Good, G. H., Ireland, P. J., Bewley, G. P., Bodenschatz, E., Collins, L. R., Warhaft, Z. 2014
Settling regimes of inertial particles in isotropic turbulence.
{\it J. Fluid Mech.} {\bf 759}, R3

\item[]
Gore, R. A., Crowe, C. T. 1989
Effect of Particle size on modulating turbulent intensity.
{\it Int. J. Multiphase Flow}
{\bf 15}, pp. 279-285

\item[]
Gualtieri, P., Picano, F., Sardina, G., Casciola, C. M. 2015
Exact regularized point particle method for multiphase flows in the two-way coupling regime.
{\it J. Fluid Mech.} {\bf 773}, pp. 520-561

\item[]
van Hinsberg, M. A. T., ten Thije Boonkkamp, J. H. M., Clercx, H. J. H. 2011
An efficient, second order method for the approximation of the Basset history force.
{\it J. Comput. Phys.} {\bf 230}, pp. 1465-1478

\item[]
Hwang, W., Eaton, J. K. 2006
Homogeneous and isotropic turbulence modulation by small heavy ($St\sim 50$) particles.
{\it J. Fluid Mech.}
{\bf 564}, pp. 361-393


\item[]
Jim\'{e}nez, J., Wray A. A., Saffman, P. G., Rogallo, R. S. 1993
The structure of intense vorticity in isotropic turbulence.
{\it J. Fluid. Mech.}
{\bf 255}, pp. 65-90

\item[]
Kajishima, T., Takiguchi, S. 2002
Interaction between particle clusters and particle-induced turbulence.
{\it Int. J. Heat Fluid Fl.}
{\bf 23}, pp. 639-646

\item[]
Kempe, T., Vowinckel, B., Fr\"{o}hlich, J. 2014
On the relevance of collision modeling for interface-resolving simulations 
of sediment transport in open channel flow.
{\it Int. J. Multiphase Flow} {\bf 58}, pp. 214-235

\item[]
Kim, J., Moin, P. 1985
Application of a fractional-step method to incompressible Navier-Stokes equations.
{\it J. Comput. Phys.}
{\bf 59}, pp. 308-323

\item[]
Kulick, J. D., Fessler, J. R., Eaton, J. K. 1994 
Particle response and turbulence modification in fully developed channel flow.
{\it J. Fluid Mech.} {\bf 277}, pp. 109-134

\item[]
Kurose, R., Komori, S. 1999
Drag and lift forces on a rotating sphere in a linear shear flow.
{\it J. Fluid. Mech.} {\bf 384}, pp. 183-206

\item[]
Lambert, R. A., Picano, F., Breugem, W. P., Brandt, L. 2013
Active suspensions in thin films: nutrient upwake and swimmer motion.
{\it J. Fluid. Mech.} {\bf 733}, pp. 528-557

\item[] 
Li, Y., Mclaughlin, J. B., Kontomaris, K., Portela, L. 2001
Numerical simulation of particle-laden turbulent channel flow.
{\it Phys. Fluids} {\bf 13}, pp. 2957-2967.

\item[]
Lucci, F., Ferrante, A., Elghobashi, S. 2010
Modulation of isotropic turbulence by particles of Taylor length-scale size.
{\it J. Fluid. Mech.}
{\bf 650}, pp. 5-55

\item[] 
Maxey, S. R., Riley, J. J. 1983
Equation of motion for a small rigid sphere in a nonuniform flow.
{\it Phys. Fluids} 
{\bf 26}, pp. 883-889.

\item[]
Maxey, M. R. 1987
The motion of small spherical particles in a cellular flow field.
{\it Phys. Fluids} {\bf 30}, pp. 1915-1928

\item[]
Mei R., Adrian, R. J. 1992
Flow past a sphere with an oscillation in the free-stream velocity and unsteady drag at finite Reynolds number. 
{\it J. Fluid Mech.} {\bf 237}, pp. 323-341

\item[]
Nielsen, P. 1993
Turbulence effects on the settling of suspended particles.
{\it J. Sedim. Petrol.} {\bf 63}, pp. 835-838

\item[]
Olivieri, S., Picano, F., Sardina, G., Iudicone, D., Brandt, L. 2014
The effect of the Basset history force on particle clustering in homogeneous and isotropic turbulence.
{\it Phys. Fluids} {\bf 26}, 041704

\item[]
Ozgoren, M. 2013
Flow structures around an equilateral triangle arrangement of three spheres.
{\it Int. J. Multiphase Flow} {\bf 53}, pp. 54-64

\item[]
Paris, A. D., Eaton, J. K. 2001
Turbulence attenuation in a particle-laden channel flow.
{\it Report TSD-137, Dept. of Mechanical Engineering, Stanford University}

\item[]
Picano, F., Breugem, W. P., Brandt, L. 2015
Turbulent channel flow of dense suspensions of neutrally buoyant spheres.
{\it J. Fluid Mech.} {\bf 764}, pp. 463-487

\item[]
Proudman, I., Pearson, J. 1957
Expansions at small Reynolds numbers for the flow past a sphere and a circular cylinder.
{\it J. Fluid Mech.} {\bf 2}, pp. 237-262

\item[]
Rani, S. L., Winkler, C. M., Vanka, S. P. 2004
Numerical simulations of turbulence modulation by dense particles in a fully developed pipe flow.
{\it Powder Technol.}
{\bf 141}, pp. 80-99

\item[]
Rubinow, S. I., Keller, J. B. 1961
The transverse force on a spinning sphere moving in a viscous fluid.
{\it J. Fluid Mech.}
{\bf 11}, pp. 447-459

\item[]
Santarelli, C.,  Fr\"{o}hlich, J. 2016
Direct Numerical Simulations of spherical bubbles in vertical turbulent
channel flow. Influence of bubble size and bidispersity.
{\it Int. J. Multiphase Flow} {\bf 81}, pp. 27-45

\item[]
Schneiders, L., Meinke, K., Schroder, W. 2017
On the accuracy of Lagrangian point-mass models for heavy non-spherical particles in isotropic turbulence.
{\it Fuel} {\bf 201}, pp. 2-14

\item[]
Squires, K. D., Eaton, J. K. 1990
Particle response and turbulence modification in isotropic turbulence.
{\it Phys. Fluids} {\bf A2}, pp. 1191-1202

\item[] 
Sridhar, G., Katz, J. 1995
Drag and lift forces on microscopic bubbles entrained by a vortex.
{\it Phys. Fluids} {\bf 7}, pp. 389-399.

\item[]
Sundaram, S., Collins, L. R., 1999
A numerical study of the modulation of isotropic turbulence by suspended particles.
{\it J. Fluid. Mech.}
{\bf 379}, pp. 105-143

\item[]
Tenneti, S., Subramaniam, S. 2014
Particle-Resolved Direct Numerical Simulation for Gas-Solid Flow Model Development. 
{\it Ann. Rev. Fluid Mech.} {\bf 46}, no. 1, pp. 199-230

\item[]
Tsuji, T., Narutomi, R., Yokomine, T., Ebara, S., Shimizu, A. 2003
Unsteady three-dimensional simulation of interactions between flow and two particles.
{\it Int. J. Multiphase Flow} {\bf 29}, pp. 1431-1450

\item[]
van Wachem, B. G. M., Schouten, J. C., van den Bleek, C. M., Krishna, R., Sinclair, J. L. 2001
Comparative Analysis of CFD Models of Dense Gas-Solid Systems.
{\it AIChE J.} {\bf 47}, pp. 1035–1051

\item[]
Wakaba L., Balachandar, S. 2005
History force on a sphere in a weak linear shear flow. 
{\it Int. J. Multiphase Flow} {\bf 31}, pp. 996-1014

\item[]
Yoon, D. H., Yang, K. S. 2007
Flow-induced forces on two nearby spheres. 
{\it Phys. Fluids} {\bf 19}, 098103

\end{itemize}

\clearpage
\pagestyle{empty}

\begin{table}[t]
  \begin{center}
    \caption{Acceleration obtained by the present estimation method.
    The subscripts 1 and 2 correspond to $x_1$ and $x_2$-components.}
    \vspace{3mm}
    \begin{tabular}{c|cc|cc|cc} \hline \rule[-17pt]{0pt}{41pt}
      $\rho_d/\rho_c$ & $a^{sv}_1$ & $a^{sv}_2$ & $a^{sv+pg}_1$ & $a^{sv+pg}_2$ & $a^{FR}_1$ & $a^{FR}_2$
      \\ \hline
      1    & $2.50$ & $-0.122$  & $2.35$  & $-18.4$  & $0.477$  & $-19.2$  \\
      10   & $-1.25$ & $-4.85$ & $-1.62$ & $-7.24$ & $-1.52$ & $-7.69$ \\
      100  & $0.292$ & $-0.314$ & $0.464$ & $-0.568$ & $0.547$ & $-0.628$ \\
      1000 & $-0.174$ & $0.174$ & $-0.176$ & $0.176$ & $-0.196$ & $0.196$ \\\hline
    \end{tabular}
    \label{tab:result_force_estimate}
  \end{center}
\end{table}

\clearpage

\clearpage
\renewcommand{\thefigure}{A\arabic{figure}}
\setcounter{figure}{0} 

\begin{figure}[!h]
\centering
\begin{overpic}[scale=0.9,angle=-0]{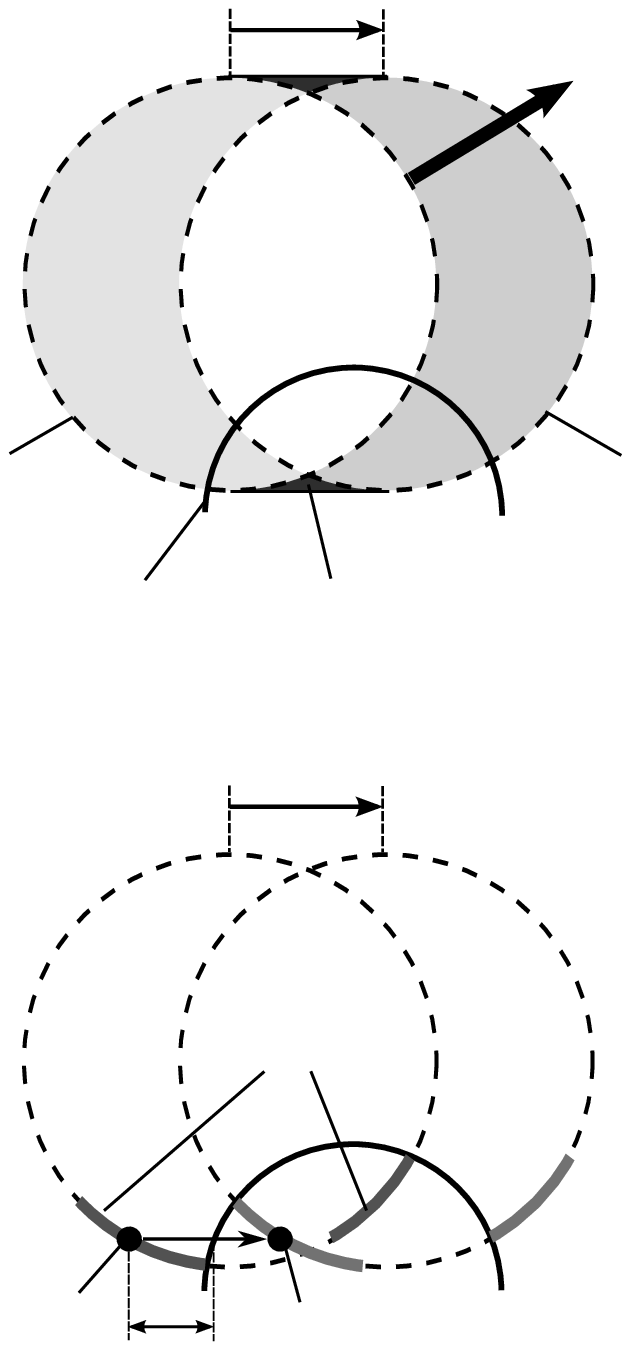}
  \put(20,99){$h\bm{e}_i$}
  \put(20,41){$h\bm{e}_i$}
  \put(43,95){$\bm{n}_V$}
  \put(-4,62){$V(\bm{x})$}
  \put(45,62){$V(\bm{x}+h\bm{e}_i)$}
  \put(22,54){$Q_{\rm cut}$}
  \put(4,54){interface}
  \put(5,78){$Q^{-}$}
  \put(35,78){$Q^{+}$}
  \put(18,22){$S_{\rm jump}$}
  \put(3,1){$\bm{S}$}
  \put(21,0){$\bm{S}+h\bm{e}_i$}
  \put(10,-2){$\beta h$}
\end{overpic}
\caption{Sketch of the geometrical difference between the volume-averaged quantities $Q(\bm{x})$ and $Q(\bm{x}+h\bm{e}_i)$, with nomenclature used in the derivations reported in Appendix A.}
\label{fig:app_diff1}
\end{figure}

\renewcommand{\thefigure}{D\arabic{figure}}
\setcounter{figure}{0} 

\begin{figure}[!h]
\centering
 \psfrag{0}[c][c][1]{$0$}
 \psfrag{x}[c][c][1]{$x$}
 \psfrag{y}[c][c][1]{$y$}
 \psfrag{a}[c][c][1]{$a$}
 \psfrag{t}[c][c][1]{$t$}
 \psfrag{r}[c][c][1]{$r_d$}
 \psfrag{R}[c][c][1]{$R$}
\begin{overpic}[scale=0.4,angle=-0]{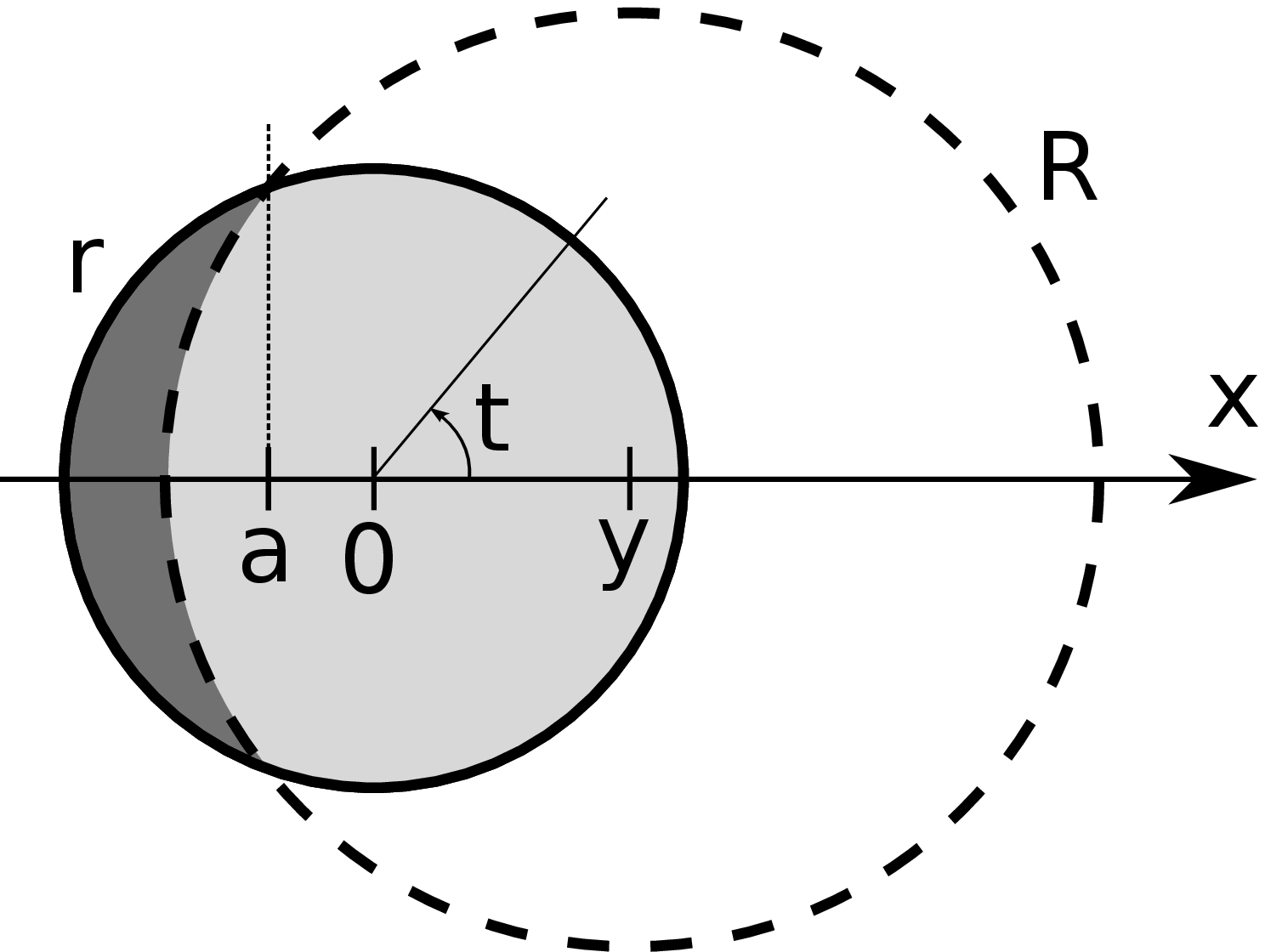}
\end{overpic}
\caption{Sketch introducing the geometrical variables used for the calculations of the volume averages.}
\label{fig:app_geom1}
\end{figure}

\renewcommand{\thefigure}{E\arabic{figure}}
\setcounter{figure}{0} 

\begin{figure}[!h]
\centering
 \psfrag{t}[s][s][1]{$t\nu/D^2$}
 \psfrag{v}[c][c][1]{$v_{p2}D/\nu$}
\begin{overpic}[scale=0.6, angle=-90]{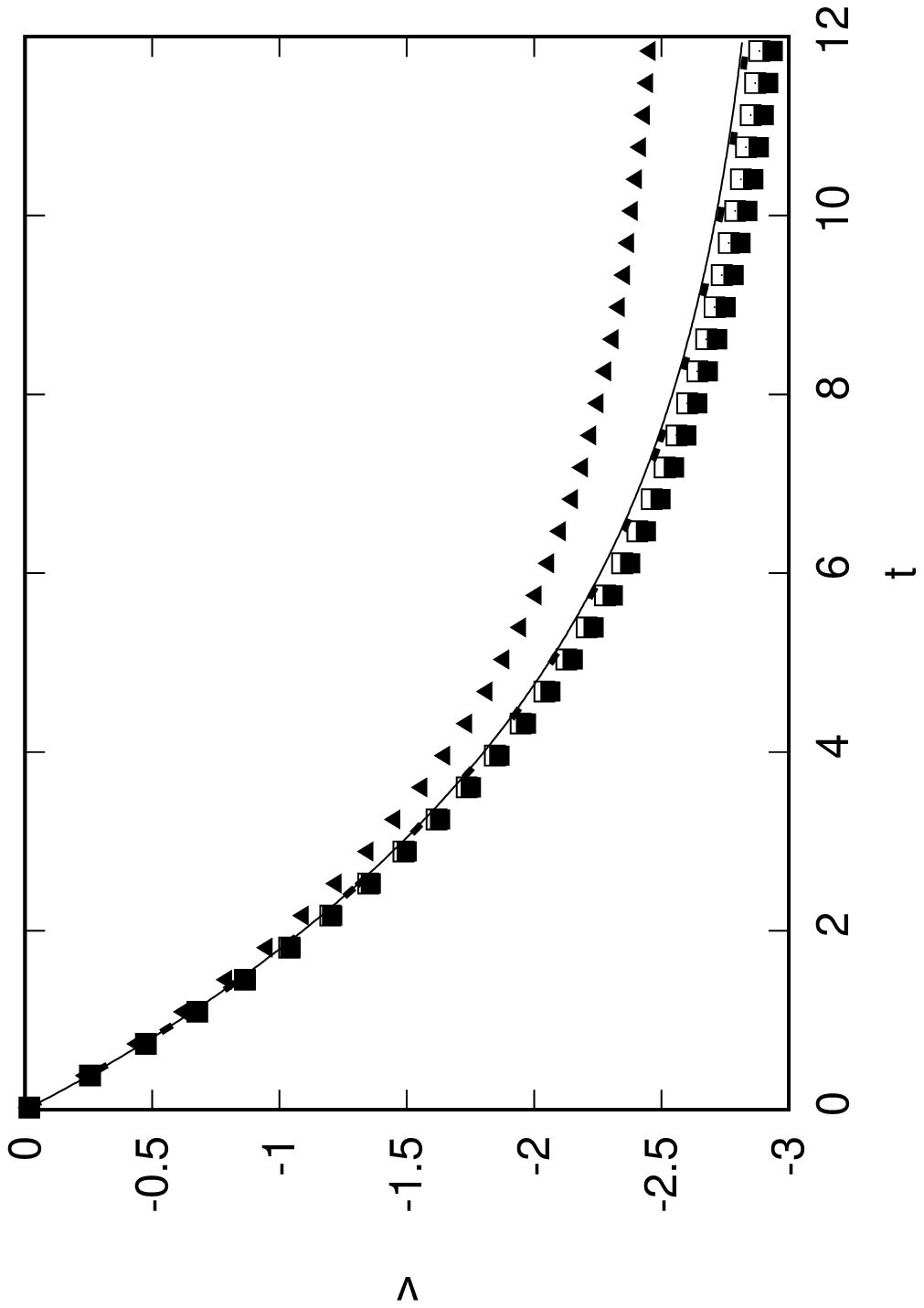}
\end{overpic}
\caption{Time evolution of the particle settling velocity in a fluid at rest.
Solid line, VA simulation ($R=0.75D$, $D/\Delta x=2$); 
dashed line, VA simulation ($R=0.75D$, $D/\Delta x=4$);
filled triangle, VA simulation ($R=0.75D$, $D/\Delta x=1$);
open square, VA simulation ($R=1.5D$, $D/\Delta x=2$);
filled square, VA simulation ($R=1.5D$, $D/\Delta x=1$).
Solid and dashed lines almost overlap with each other.}
\label{fig:app_falling_vel_R1.5}
\end{figure}

\begin{figure}[!h]
\centering
 \psfrag{x}[s][s][1]{$x_1/\pi L$}
 \psfrag{y}[c][c][1]{$x_2/\pi L$}
\begin{overpic}[scale=0.7, angle=-90]{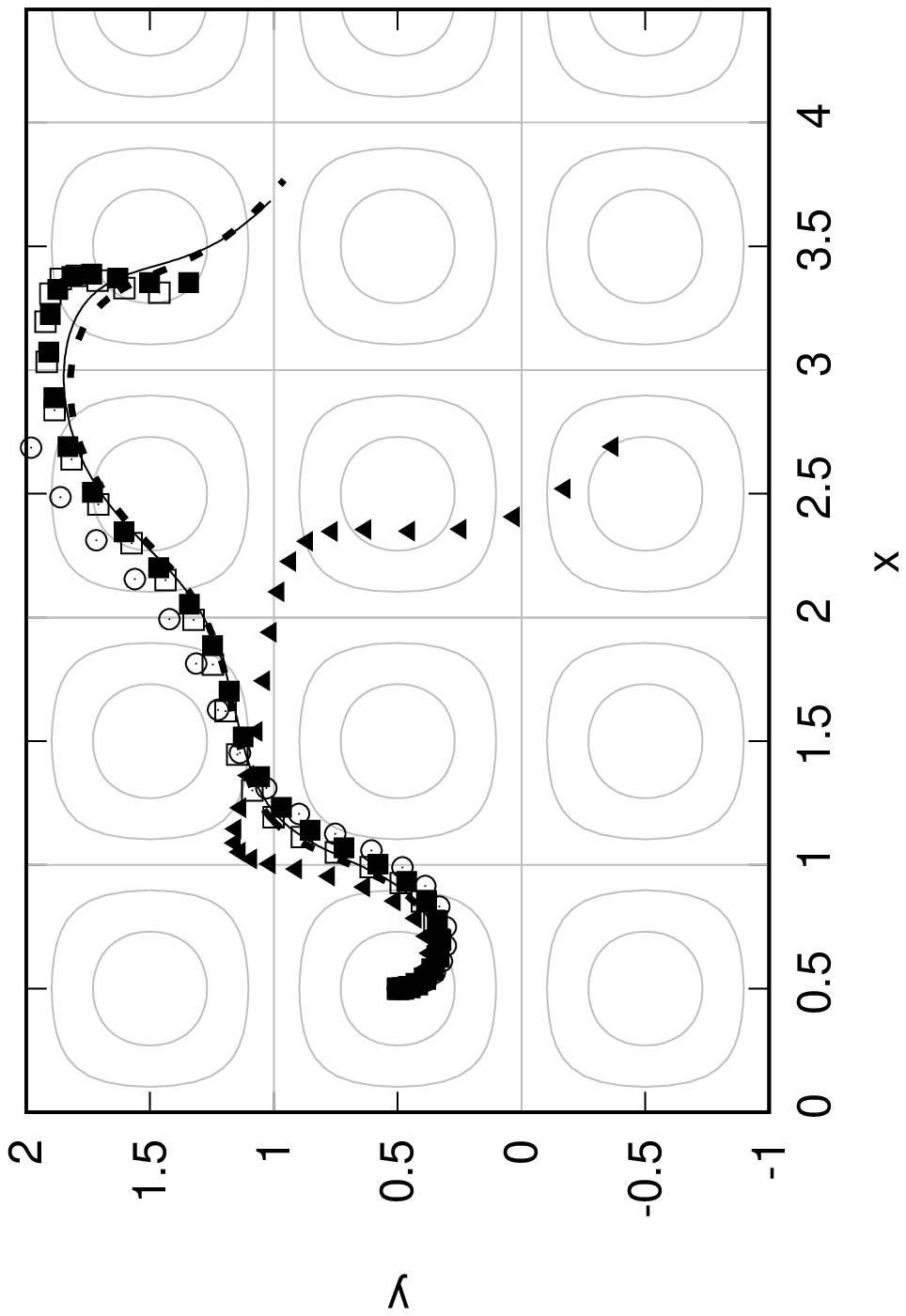}
\end{overpic}
\caption{Trajectory of a single particle with density ratio $\rho_d/\rho_c=100$ under the gravity in an array of Taylor-Green vortices.
Solid line, VA simulation ($R=0.75D$, $D/\Delta x=2$); 
dashed line, VA simulation ($R=0.75D$, $D/\Delta x=4$);
filled triangle, VA simulation ($R=0.75D$, $D/\Delta x=1$);
open square, VA simulation ($R=1.5D$, $D/\Delta x=2$);
filled square, VA simulation ($R=1.5D$, $D/\Delta x=1$);
open circle, O-NL simulation.
The grey lines show the streamlines of the undisturbed flow.}
\label{fig:app_celljob_test_xy_R1.5}
\end{figure}

\end{document}